\documentclass[journal]{IEEEtran}
\usepackage{color}
\usepackage{graphicx}
\usepackage{amsmath}
\usepackage{amsfonts, psfrag}
\usepackage{amssymb}
\usepackage{booktabs}
\usepackage{cases}
\usepackage{tabularx}
\usepackage{boxedminipage}
\usepackage{url}

\begin{document}

% paper title
\title{Asymptotically Optimal Power Allocation for Energy Harvesting Communication Networks}
\author{ Nikola Zlatanov,~\IEEEmembership{Member,~IEEE,}    Robert Schober,~\IEEEmembership{Fellow,~IEEE,} and Zoran Hadzi-Velkov,~\IEEEmembership{Senior Member,~IEEE} 
\thanks{Manuscript received November 16, 2015; revised July 19, 2016; accepted
January 30, 2017.  This work has been presented  in part at  IEEE Globecom, Atlanta, Dec. 2013 \cite{6831450}.}
\thanks{N. Zlatanov is  with the Department of Electrical and Computer Systems Engineering, Monash University, Melbourne, VIC 3800, Australia (e-mail: nikola.zlatanov@monash.edu).}
\thanks{R. Schober is with the Friedrich-Alexander University of Erlangen-N\"urnberg,
Institute for Digital Communications, D-91058 Erlangen, Germany
(e-mail: robert.schober@fau.de ).}
\thanks{Z. Hadzi-Velkov  is with the Faculty of Electrical Engineering and Information Technologies, Sts. Cyril and Methodius University, 1000 Skopje, Macedonia (e-mail: zoranhv@feit.ukim.edu.mk).
\newline}
%\thanks{Digital Object Identif\-ier 10}
%\vspace*{-12mm}
 } 
%\markboth{Revision submitted for possible publication in the
%IEEE Transactions on Information Theory}{Shell
%\MakeLowercase{\textit{et al.}}: Bare Demo of IEEEtran.cls for
%Journals}

\maketitle

\begin{abstract}
For a general energy harvesting (EH) communication network, i.e., a network where the nodes generate their transmit power through EH, we derive the asymptotically optimal online power allocation solution   which optimizes  a general utility function  when the number of transmit time slots,  $N$,  and the battery capacities of the EH nodes,  $B_{\rm max}$,   satisfy $N\to\infty$ and  $B_{\rm max}\to\infty$.
  The considered family of utility functions is  general enough to include the most important performance measures in communication theory such as the average data rate, outage probability, average bit error probability, and average signal-to-noise ratio.  The proposed power allocation solution is very simple. Namely, the asymptotically  optimal power allocation for the EH network is identical to the optimal power allocation for an equivalent  non-EH network whose nodes have infinite energy available but their average transmit power is constrained to be equal to the average harvested power  and/or the maximum average transmit power of the corresponding nodes in the EH network. Moreover, the maximum average performance of a general EH network  converges to the maximum average performance of the corresponding equivalent non-EH network, when  $N\to\infty$ and  $B_{\rm max}\to\infty$.  
 Although the proposed solution is asymptotic in nature,   it is applicable to EH systems transmitting in a large but finite number of  time slots and having a  battery capacity  much larger than the  average harvested power  and/or the maximum   average transmit power.
\end{abstract}

\IEEEpeerreviewmaketitle

\newtheorem{theorem}{Theorem}
\newtheorem{corollary}{Corollary}
\newtheorem{remark}{Remark}
\newtheorem{lemma}{Lemma}
\newtheorem{defi}{Definition}
\newtheorem{proposition}{Proposition}
\thispagestyle{empty}

%%%%%%%%%%%%%%%%%%%%%%%%%%%%%%%%%%%%%%%%%%%%%%%%%%%%%%%%%%%%%%%%%
\section{Introduction}
Energy harvesting (EH) transmitters  collect random amounts of energy and store them in their batteries. For this purpose, several techniques for harvesting energy from various renewable sources, such as pressure, motion, solar, etc.,  have  been proposed, see \cite{ottman2002adaptive, mitcheson2008energy, paradiso2005energy}, and  references therein.  Using the stored harvested energy,   EH transmitters  send codewords (or uncoded symbols)  to their designated receivers. To this end,   the codewords' powers have to be adapted to the random amounts of harvested energy  and  also to the quality of the channels between the EH transmitters and their designated receivers, which may be time-varying  due to fading.  Excellent  overviews of recent advances in EH technology are provided in \cite{7010878}, \cite{7081085}, \cite{kuadvances}.

In the literature, there are two approaches for solving the EH power allocation problem. The aim of the first approach is to obtain the optimal \textit{online} solution. Online solutions require  only   causal energy and channel  state information (CSI), therefore, they are  feasible in practice. However, for finite numbers of transmit time slots, the optimal online solution  often  cannot be computed even for simple communication channels, such as the point-to-point channel. This is due to the fact that computing the optimal online solution typically involves dynamic programing \cite{ho2012optimal}, \cite{ozel2011transmission}. The computational complexity of dynamic  programing, even for  the simple point-to-point channel,  grows exponentially with the number of transmit time slots, and therefore, cannot be computed even for small-to-moderate numbers of codewords \cite{ho2012optimal}.
As a result, the second approach whose objective  is to obtain the optimal \textit{offline}  solution is often adopted in the literature \cite{ho2012optimal}, \cite{ozel2011transmission}.  Offline solutions require   non-causal energy and CSI, therefore, they are  not feasible in practice. Nevertheless, offline solutions  may still serve as  performance upper bounds for the performance of any online solution. 
\nocite{ho2012optimal,  ozel2011transmission, 5441354, 6108309, 7001259, 5992840, 6177987, 6135979, 6253062, 7010000, 6214501, 6381384, 5585638, 6807761, 6425377, 6307793} 

In the literature, in general,  the optimal offline solution is studied for a specific system model, e.g., the point-to-point channel, the broadcast channel, etc., and   a specific performance measure, most often the achievable data rate  \cite{ho2012optimal}-\cite{6307793} and seldom other performance measures  such as the outage probability \cite{6697931}, \cite{7009991}. Hence,  the  proposed solutions  and  the  framework  for deriving these  solutions  are usually applicable to the specific considered system model and the specific considered performance measure only, and cannot be easily generalized to different system models and/or  different performance measures. For example, the optimal offline power allocation which maximizes the achievable data rate     has been investigated for the point-to-point channel  in \cite{ho2012optimal}-\cite{5441354}, for the broadcast channel in  \cite{6108309}-\cite{6135979}, for the multiple-access channel  in    \cite{6253062}-\cite{6214501}, and for the  relay channel in \cite{6381384}-\cite{6307793}. The outage probability for the  point-to-point EH channel has been investigated in \cite{6697931} and \cite{7009991}.
 The above references make different  assumptions about the battery capacities and the numbers of transmit time slots, namely, they assume finite and/or infinite battery capacities and  finite and/or infinite numbers of transmit time slots.  
On the other hand, in the cases where  optimal online solutions are provided, the solutions are based on dynamic programing and thus, can not be computed  even for small-to-moderate  numbers of transmit time slots, see  for example \cite{ho2012optimal}, or they are derived for infinite numbers of transmit time slots and infinite battery capacities and are applicable to a specific  model and specific performance measure only, see  for example \cite{6216430}.    

Hence, as seen from the discussion above, optimal online solutions \textit{for a general EH network}  which  maximize some average performance measure, such as the average data rate, the outage probability, the average bit error probability, and the average signal-to-noise ratio (SNR),   are not known.  
Motivated by this, the objective of this paper is to  develop a framework for obtaining the optimal online power allocation solution which   maximizes some   predefined utility function  of a \textit{general EH communication network}.
 The admissible utility functions are  general enough to include the most important average performance measures in communication theory, including the average data rate, outage probability, average bit error probability, and average SNR. The developed  framework is asymptotic and holds when the number of transmit time slots, $N$,   and the battery capacity, $B_{\max}$, at each EH node in the network are   infinite. Based on the developed  framework the optimal online solution  for a general EH network with a general utility function   is relatively easy to obtain. Namely, the optimal online power allocation for the EH network is given by the optimal online power allocation for an equivalent non-EH network where each node  has  infinite available energy, under the constraint that each node in the non-EH network employs the same average transmit power as the corresponding node in the EH network. As a consequence, the maximum average performance of the EH network  converges to the maximum average performance of the corresponding equivalent non-EH network, as $N\to\infty$ and $B_{\max}\to\infty$. 
Hence, the  EH network suffers no average performance loss compared to the equivalent non-EH network. Therefore, in the asymptotic case, instead of finding the optimal power allocation for the EH network, it is sufficient to  find  the optimal power allocation for the corresponding equivalent  non-EH network and apply it to the EH network.

The practical value of the developed framework is that it gives an average performance upper bound for any online power allocation  in general EH networks with finite $N$ and/or $B_{\rm max}$. Furthermore, in practice, the proposed online solution is applicable to  EH networks transmitting in a large but finite number  of time slots and having nodes with battery capacities  much larger than their  average harvested powers  and/or   the maximum  average transmit powers.

In order to  introduce the proposed framework for general EH networks step-by-step, we   first present a corresponding  framework for the point-to-point EH system in Section  II. Then, we generalize the framework to the broadcast and  multiple-access EH networks in Section III. Finally, in Section IV, we further generalize the framework to  general EH networks.  In Section V,  we illustrate the applicability of the developed framework through  numerical examples, and  Section VI concludes the paper.

\section{The Point-to-Point EH System}

In the following, we consider the point-to-point EH communication system and formulate  the corresponding power allocation problem. Then, we define an equivalent point-to-point non-EH system which differs from the EH  system only  in the energy  available  for transmission of the codewords. For both systems, the transmission time is divided into slots of equal length, each codeword\footnote{For the case when one symbol spans one time slot and the power allocation has to be performed in a symbol-by-symbol manner, we will replace the word ''codeword" by the word ''symbol".} spans one time slot, and the power allocation has to be performed in a slot-by-slot (i.e., codeword-by-codeword) manner.   Furthermore, the  number of transmit time slots, $N$, satisfies $N\to\infty$. We note that all of the assumptions and definitions that we introduce for the EH transmitter in the point-to-point EH system are also valid for the individual EH transmitters in the more complex EH networks considered in Sections III and IV.

\subsection{Point-to-Point EH System Model}\label{s-eh}
We consider an EH transmitter which harvests random amounts of energy in each time slot and stores them in its battery.  It uses the  energy stored   in its battery to transmit codewords to a receiver. 
Let the  capacity of the battery, denoted by $B_{\rm max}$,  be unlimited, i.e., $B_{\rm max}\to\infty$ holds. Let $B(i)$ denote the amount of power\footnote{In this paper, we adopt the normalized energy unit  Joule-per-second. As  a result, we use  the terms "energy" and "power"  interchangeably.}  available
 in the battery at the end of time slot $i$. Let the amount of harvested power that is added to the battery storage in time slot $i$  be denoted by $P_{\rm in}(i)$. We assume that $P_{\rm in}(i)$ is a stationary and ergodic random process with average  $\bar P_{\rm in}$ given by
\begin{eqnarray}\label{eq_P_in}
\bar P_{\rm in}=\lim_{N\to\infty}\frac{1}{N}\sum_{i=1}^N P_{\rm in}(i) =E\{P_{\rm in}(i)\},
\end{eqnarray}
where $E\{\cdot\}$ denotes expectation, and   the converges of the mean in (\ref{eq_P_in})  is almost surely.

In order to formulate the optimal power allocation solution, in the following, we introduce the desired amount of  power that the EH transmitter wants to extract from the battery  in time slot $i$, denoted by $P_{\rm d}(i)$, which satisfies $0 \leq P_{\rm d}(i)<\infty$, $\forall i$. Note that, in contrast  to a non-EH system, in  an EH system, the desired amount of power that we want to extract from the battery in time slot $i$   and the actual amount of power that can be extracted from the battery in time slot $i$ may not be identical. In particular, we may desire more power than what is currently available in the battery. Let  $P_{\rm out}(i)$ denote the  actual amount of  power extracted from the battery   in time slot $i$  and used for  transmission of the $i$-th codeword. Then, the relation between $P_{\rm d}(i)$ and $P_{\rm out}(i)$ is given by
\begin{eqnarray}\label{eq_p_pout}
    P_{\rm out}(i)=\min\{B(i-1),P_{\rm d}(i)\},
\end{eqnarray}
i.e., the power  extracted from the battery at time slot $i$, $P_{\rm out}(i)$,  is limited by the desired amount of power that the EH transmitter wants to extract from the battery,  $P_{\rm d}(i)$,  and the amount of power  stored in the battery at the end of the previous time slot, $B(i-1)$. Obviously, $P_{\rm out}(i)\leq P_{\rm d}(i)$, $\forall i$, always holds. Considering the  harvested power, $P_{\rm in}(i)$, and the extracted power, $P_{\rm out}(i)$,    the amount of power stored in the battery at the end of time slot $i$ is given by
\begin{eqnarray}\label{eq_q}
    B(i)=B(i-1)+P_{\rm in}(i)-P_{\rm out}(i).
\end{eqnarray}
Since $P_{\rm in}(i)$ is a stationary and  ergodic random process, and  as a result of   the law of conservation of flow in the battery, the time average of  $P_{\rm out}(i)$ converges to a finite number, which we call the average transmit power,  denoted by  $\bar P_{\rm out}$,   given by
\begin{align}\label{eq_P_out_av}
    \bar P_{\rm out}&=\lim_{N\to\infty}\frac{1}{N}\sum_{i=1}^N P_{\rm out}(i)\nonumber\\
 & =\lim_{N\to\infty}\frac{1}{N}\sum_{i=1}^N \min\{B(i-1), P_{\rm d}(i)\}.
\end{align}
Note that $\bar P_{\rm out}$ depends on $P_{\rm in}(i)$ and $P_{\rm d}(i)$, $\forall i$.
On the other hand, the average desired power that the  EH transmitter wants to extract from the battery,  denoted by  $\bar P_{\rm d}$, is given by 
\begin{align}\label{eq_asds}
 \bar P_{\rm d}=\lim\limits_{N\to\infty} \frac{1}{N} \sum_{i=1}^N P_{\rm d}(i), 
\end{align}
where we assume  that the $P_{\rm d}(i)$, $\forall i$, are such that the limit in (\ref{eq_asds}) holds. 
 Note that $\bar P_{\rm out}\leq \bar P_{\rm d}$ as a result of (\ref{eq_P_out_av}).

In communication systems, typically constraints are imposed on the transmit power $P_{\rm out}(i)$ and/or the average transmit power  $\bar P_{\rm out}$. Considering these constraints, let $\bar P_{\rm lim}$ denote the
 upper limit on the average transmit power $\bar P_{\rm out}$ for the EH transmitter such that $ \bar P_{\rm out}\leq \bar P_{\rm lim}$ has to hold. Note that if there are no  constraints imposed  on   $P_{\rm out}(i)$ and $\bar P_{\rm out}$, then $\bar P_{\rm lim}$ can be set to $\bar P_{\rm lim}=\infty$. We now introduce the considered class of utility functions.

\begin{defi}\label{def_1}
The utility function, denoted by $U(i)$, is associated with the $i$-th codeword. It is a predefined function that  measures some desired quality of the $i$-th codeword.   Furthermore, the considered  $U(i)$ have the following properties.

\begin{itemize}
\item[1.] $U(i)$  depends on the transmit power $P_{\rm out}(i)$. To emphasize this dependence we use the notation   $U(P_{\rm out}(i))$. 

\item[2.]  $U(P_{\rm out}(i))$ is finite for finite $P_{\rm out}(i)$, i.e., $|U(P_{\rm out}(i))|<\infty$ for $P_{\rm out}(i)<\infty$.

\item[3.]   The time average of $U(P_{\rm out}(i))$  exists, it has a finite value denoted by $\bar U$, and is given by
\begin{eqnarray}\label{av_ut_fun}
 \bar U=\lim_{N\to\infty}\frac{1}{N}\sum_{i=1}^N U(P_{\rm out}(i)).
\end{eqnarray}

\item[4.]   If we add  a negligible amount of power $\epsilon(i)>0$ to $P_{\rm out}(i)$,  $\forall i$, where $
 \lim\limits_{N\to\infty}\frac{1}{N}\sum_{i=1}^N \epsilon(i) =0 $
holds, then $\bar U$   satisfies
\begin{align}\label{av_ut_fufgn}
 \bar U&=\lim_{N\to\infty}\frac{1}{N}\sum_{i=1}^N U(P_{\rm out}(i)+\epsilon(i)) 
\nonumber\\
&= \lim_{N\to\infty}\frac{1}{N}\sum_{i=1}^N U(P_{\rm out}(i)) ,
\end{align}
i.e., adding   zero average  power cannot have a non-negligible effect on   $\bar U$.

\item[5.]  The maximum of $\bar U$ cannot decrease if the average transmit power $\bar P_{\rm out}$ increases. More precisely,
\begin{align}\label{eq_1}
 &\max\limits_{P_{\rm out,1}(i)} \lim_{N\to\infty}\frac{1}{N}\sum_{i=1}^N U(P_{\rm out,1}(i)) \nonumber\\
&\leq  \max\limits_{P_{\rm out,2}(i)} \lim_{N\to\infty}\frac{1}{N}\sum_{i=1}^N U(P_{\rm out,2}(i))
\end{align}
holds for 
\begin{align}\label{eq_1a}
   \lim_{N\to\infty}\frac{1}{N}\sum_{i=1}^N  P_{\rm out,1}(i) \leq   \lim_{N\to\infty}\frac{1}{N}\sum_{i=1}^N  P_{\rm out,2}(i).
\end{align}
 
\end{itemize}
\end{defi}
\begin{remark}
Given the above properties of the utility function $U(P_{\rm out}(i))$, valid utility functions include the data rate   of the $i$-th codeword, the average  SNR  of the  $i$-th codeword at the receiver, the outage probability of the   $i$-th codeword, and the symbol  (bit) error probability of the $i$-th symbol (bit) for uncoded transmission. Hence, our definition of $U(P_{\rm out}(i))$ includes the most  important performance  metrics  in communication theory. 
\end{remark}

\begin{remark}\label{rem_2}
For simplicity of notation, we have assumed that $\bar U$ is maximized. However, similar results can be obtained if $\bar U$ needs to be minimized.
\end{remark}

Considering how we have modeled the powers in the EH system,  the desired powers $P_{\rm d}(i)$, $\forall i$, are the only variables with a degree of freedom, since $P_{\rm out}(i)$ is a function of $P_{\rm in}(j)$ and  $P_{\rm d}(j)$, for $j=1,...,i$. Given a limit on the average transmit power, $\bar P_{\rm lim}$, we want to devise an optimal power allocation strategy that maximizes the average utility function $\bar  U$. More precisely,  we want to determine the optimal desired powers $P_{\rm d}(i)\in \mathcal{P}$, $\forall i$, where $\mathcal{P}$ is the domain of $P_{\rm d}(i)$  given by  $0\leq P_{\rm d}(i)<\infty$,
 which produce a corresponding $P_{\rm out}(i)$, $\forall i$,  such that $\bar P_{\rm out}\leq \bar  P_{\rm lim}$ holds and the average utility function  $\bar U$  is maximized.  We state this rigorously in the following maximization problem: 
\begin{eqnarray}\label{MPR1}
\begin{array}{ll}
 {\underset{P_{\rm d}(i)\in\mathcal{P},\forall i}{\rm{Maximize: }}}&\lim\limits_{N\to\infty}\frac{1}{N}\sum_{i=1}^N U(i)\\
{\rm{Subject\;\; to: }} & {\rm C1:}\;   P_{\rm out}(i)= \min\{B(i-1),P_{\rm d}(i)\} \\
& {\rm C2:}\; \bar P_{\rm out}\leq \bar P_{\rm lim} \\
& {\rm C3:}\; \textrm {Optional  constraints on }  P_{\rm d}(i)\\
& {\rm C4:}\; B(i)=B(i-1)+P_{\rm in}(i)-P_{\rm out}(i),\;\;
\end{array}
\hspace{-5mm}
\end{eqnarray} 
where we assume that $P_{\rm in}(i)$ is known causally at the EH transmitter. More precisely,   the amount of harvested power during the $i$-th time slot is revealed at the EH transmitter at the end of the $i$-th time slot. Furthermore, C3 represents optional   constraints on $P_{\rm d}(i)$, if any. For example, C3 may constrain $P_{\rm d}(i)$  to be constant for all time slots, or not to exceed some upper limit, or to be zero in certain time slots.  We assume that the constraints on $P_{\rm d}(i)$, $\forall i$, if they exist, are such that they allow $\bar P_{\rm d}=\bar P_{\rm \lim}$ to be achievable. Otherwise, if $\bar P_{\rm d}=\bar P_{\rm \lim}$ is not achievable, then instead of  $\bar P_{\rm \lim}$ in (\ref{MPR1}) we can introduce another upper limit on the average transmit power, denoted by  $\bar{ P}_{\rm \lim, new}$, which is the maximum possible $\bar P_{\rm d}$ allowed by C3. Then, $\bar P_{\rm d}=\bar P_{\rm \lim, new}$ is achievable and we just need to replace  $\bar{ P}_{\rm \lim}$ in C3   with $\bar{ P}_{\rm \lim, new}$\footnote{We note that the assumption that $\bar P_{\rm d}=\bar P_{\rm lim}$ is achievable will be used for all individual EH transmitters considered in this paper.}. 

\begin{remark}
Note that there is a difference between $\bar P_{\rm d}$ and $\bar P_{\rm lim}$. $\bar P_{\rm d}$ is a design parameter of the communication system that the designer can choose and optimize such that the optimal power allocation  is obtained. On the other hand, $\bar P_{\rm lim}$ is a constraint of the communication system that the system designer cannot influence. Instead, the system designer has to optimize the power allocation   such that the constraint   $\bar P_{\rm out}\leq \bar P_{\rm lim}$ is  satisfied.
\end{remark}

Our objective is to solve (\ref{MPR1}), i.e., to obtain the desired powers $P_{\rm d}(i)\in \mathcal{P}$, $\forall i$, which produce  transmit powers $P_{\rm out}(i)$, $\forall i$, that  maximize the average utility function $\bar U$ and satisfy all of the corresponding constraints. To this end, we introduce a non-EH communication system  having  infinite energy  available for  transmission of the codewords that is equivalent to the EH system. This non-EH system is defined in the following subsection.

\subsection{Equivalent Point-to-Point  non-EH  System}
The equivalent  point-to-point non-EH system is   identical   to the EH  system, defined in Section~\ref{s-eh}, but with the following two differences. First, the non-EH system has infinite\footnote{The assumption of the availability of infinite amounts of power is needed for mathematical convenience. In practice, the power available in non-EH systems is limited as well, of course.} power available   for the transmission of each of its codewords, and secondly, the upper limit on its average transmit power is $\bar P_{\rm lim,non-EH}$. We will  show  that if $\bar P_{\rm lim,non-EH}$ is appropriately adjusted, the EH system and the non-EH system become equivalent in terms of maximum average performance, cf. Theorem 1, when $N\to\infty$. As a result of the infinite  power available  in the non-EH system, any   desired  power  $ P_{\rm d}(i) \in \mathcal{P}$  can be provided and  therefore  the transmit power of each codeword,  $P_{\rm out}(i)$, is identical to the desired power, i.e., $P_{\rm out}(i)=P_{\rm d}(i)$, $\forall i$, holds. This is the fundamental difference between an EH and a non-EH system since, contrary to a non-EH system, in an   EH system not every desired power $P_{\rm d}(i)\in \mathcal{P}$ can be provided   by  the power supply (i.e., battery) and therefore the transmit power of the $i$-th codeword is given by (\ref{eq_p_pout}).

For the equivalent non-EH system, the aim is again to maximize the average utility function, given an upper limit on the average transmit power. However, since in this case $P_{\rm d}(i)=P_{\rm out}(i)$, $\forall i$, the  maximization problem is given by 
\begin{eqnarray}\label{MPR1_non-EH}
\begin{array}{ll}
 {\underset{ P_{\rm d}(i)\in\mathcal{P}, \forall i}{\rm{Maximize: }}}&\lim\limits_{N\to\infty}\frac{1}{N}\sum_{i=1}^N U(i)\\
{\rm{Subject\;\; to: }} &     {\rm C1:}\; P_{\rm out}(i)=P_{\rm d}(i)\\
 &  {\rm C2:} \bar P_{\rm d}\leq \bar P_{\rm lim,non-EH} \\
& {\rm C3:}\; \textrm {Optional constraints on }  P_{\rm d}(i),
\end{array}
\end{eqnarray} 
where $U(i)$, $\forall i$,  and  C3 are as in (\ref{MPR1}).
The  optimization problem in (\ref{MPR1_non-EH}) is the conventional power allocation  problem for a conventional (non-EH) point-to-point communication system, and  has been solved in the literature for many different utility functions, e.g., \cite{goldsmith1997capacity, 771147}. 

\begin{remark}\label{rem_conv1}
Note that the optimization problems in (\ref{MPR1}) and (\ref{MPR1_non-EH}) may be non-concave and  difficult to solve in general.
\end{remark}

In the following, we provide the framework for solving (\ref{MPR1}). In particular, we show that, for $N\to\infty$ and $B_{\rm max}\to\infty$, the EH optimization problem in (\ref{MPR1})  becomes   identical to the non-EH optimization problem  in  (\ref{MPR1_non-EH}), if $\bar P_{\rm lim,non-EH}$ in (\ref{MPR1_non-EH}) is appropriately adjusted. Therefore, the optimized average performance of the EH system becomes identical to the optimized average performance of the equivalent non-EH system with adjusted $\bar P_{\rm lim,non-EH}$. 
As a result, the solution of the non-EH optimization problem  is also the solution of the EH optimization problem. In other words, instead of solving the EH optimization problem in (\ref{MPR1}), one only needs to solve the non-EH optimization problem in (\ref{MPR1_non-EH}) and apply the solution  in the EH system. This is the subject of the following subsection.

\subsection{Asymptotically Optimal Power Allocation for the Point-to-Point EH System}

Before providing the solution of the EH optimization problem in (\ref{MPR1}), we introduce the following definition. 
\begin{defi}
When we say that $P_{\rm out}(i)=P_{\rm d}(i)$ holds for \textit{practically all time slots}, we mean that it holds  for all $N\to\infty$ time slots except for a negligible fraction of them, denoted by $\Delta$, which satisfies  $\lim\limits_{N\to\infty} \Delta/N=0$. 
\end{defi}

 useful lemma.
\begin{lemma}\label{lemma_1}
In the point-to-point  EH system, if $P_{\rm d}(i)$, $\forall i$, are chosen such that they   satisfy the following constraints
\begin{subequations}\label{eq_lem_1}
\begin{align}
\textbf{C1: }& 0 \leq P_{\rm d}(i)<\infty,\;\; \forall i,\\
\textbf{C2: }& \lim_{N\to\infty}\frac{1}{N}\sum_{i=1}^N P_{\rm d}(i)=  
\bar P_{\rm d}=\min\{\bar P_{\lim}, \bar P_{\rm in}\}\label{eq_lem_1b},
\end{align}
\end{subequations}
then $P_{\rm out}(i)=P_{\rm d}(i)$ will hold for practically all time slots  $i$.  Moreover, when the constraints in (\ref{eq_lem_1}) hold, the events for which  $P_{\rm out}(i) \neq P_{\rm d}(i)$  holds  have negligible contribution to the average performance, $\bar U$, when $N\to\infty$ and $B_{\rm max}\to\infty$, and therefore these events can be neglected. Thereby, by choosing  the values of $P_{\rm d}(i)$, $\forall i$, freely, as long as the constraints in (\ref{eq_lem_1}) hold, we actually choose the  values of $P_{\rm out}(i)$  freely   for practically all time slots $i$.
\end{lemma}
\begin{IEEEproof}
Please refer to Appendix~\ref{app_1}.
\end{IEEEproof}

Using, Lemma~\ref{lemma_1}, we now provide the asymptotically optimal power allocation for the point-to-point EH system  in the following theorem.
\begin{theorem}\label{theo_1}
The solution of the non-EH optimization problem in (\ref{MPR1_non-EH}) with $\bar P_{\rm lim,non-EH}$ set as $\bar P_{\rm lim,non-EH}=\min\{\bar P_{\rm lim}, \bar P_{\rm in}\}$ is also the solution to the EH optimization problem in (\ref{MPR1}). As a result,  the maximum average performance of a point-to-point  EH system, $\bar U$, is identical to the maximum average performance of its equivalent non-EH system with $\bar P_{\rm lim,non-EH}=\min\{\bar P_{\rm lim}, \bar P_{\rm in}\}$. 
\end{theorem} 
\begin{IEEEproof}
In the optimization problem in (\ref{MPR1}), we add the constraint in  (\ref{eq_lem_1b}). 
As a result, we obtain a new optimization problem for the  point-to-point EH  system where the constraints in (\ref{eq_lem_1}) are satisfied. Now, according to Lemma~\ref{lemma_1}, since the constraints in (\ref{eq_lem_1}) hold,   $P_{\rm out}(i) = P_{\rm d}(i)$ holds practically always. Hence,   we can write constraint C1 in  (\ref{MPR1})  as $P_{\rm out}(i)=  P_{\rm d}(i)$ and constraint C2 in (\ref{MPR1})  as $\bar  P_{\rm out} = \bar  P_{\rm d}=\min\{\bar P_{\rm in},\bar P_{\lim}\} $. Consequently, constraints   C4  in  the EH optimization problem in (\ref{MPR1}) becomes unnecessary, and therefore the optimization problem in (\ref{MPR1}) becomes identical to the optimization problem in (\ref{MPR1_non-EH}) with $\bar P_{\rm lim,non-EH}=\min\{\bar P_{\rm in},\bar P_{\lim}\} $. This completes the proof  of Theorem~\ref{theo_1}.
\end{IEEEproof}

Theorem~\ref{theo_1} gives a very simple solution to the power allocation problem for the  point-to-point EH system when $N\to\infty$ and $B_{\max}\to\infty$. It states that, instead of solving the  power allocation problem for the point-to-point  EH system, we should solve the power allocation problem for its equivalent   point-to-point non-EH system with $\bar P_{\rm lim,non-EH}=\min\{\bar P_{\rm lim}, \bar P_{\rm in}\}$. Then, the derived  solution for $P_{\rm d}(i)$ obtained for the equivalent non-EH system is also the solution  for  the point-to-point  EH  system. Moreover, with this solution for $P_{\rm d}(i)$, both the EH and non-EH systems achieve  the same maximum average performance $\bar U$.   The  convergence of the maximum average performance, $\bar U$, of the point-to-point EH system to the maximum average performance of its equivalent  non-EH system is a result of $P_{\rm out}(i)=P_{\rm d}(i)$ holding for practically all time slots $i$.

\begin{remark}\label{rem_4}
An interesting consequence arising from Theorem~\ref{theo_1} is that the asymptotically optimal  power allocation for the EH system  requires  only knowledge  of the average harvested power  $\bar P_{\rm in}$ and does not need  any  causal or noncausal knowledge of $P_{\rm in}(i)$, $\forall i$, i.e., any additional knowledge would not increase $\bar U$. We note that since $P_{\rm in}(i)$ is a stationary and ergodic random process, its mean, $\bar P_{\rm in}$, can be estimated from its samples. For example, an estimate of the mean harvested power in time slot $i$, denoted by $\bar P_{\rm in}^{\epsilon}(i)$, can be obtained as
\begin{align}\label{eq_sdfdxs}
\bar P_{\rm in}^{\epsilon}(i)=  \frac{i-1}{i}\bar P_{\rm in}^{\epsilon}(i-1)+ \frac{1}{i} P_{\rm in}(i).
\end{align}
Using the above recursive equation,  $\bar P_{\rm in}^{\epsilon}(i)$   approaches $ \bar P_{\rm in}$, i.e.,  $\bar P_{\rm in}^{\epsilon}(i)\to  \bar P_{\rm in}$, as  $i\to\infty$.
\end{remark}

In the following, we present several examples for the applicability of the proposed framework for the  point-to-point EH system. 
 
%%%%%%%%%%%%%%%%%%%%%%%%%%%%%%%%%%%%%%%%%%%%%%%%%%

\subsection{Examples for Power Allocation in Point-to-Point EH Systems with Fading}
In this subsection, we illustrate the application of the proposed framework, for an EH system with fading. To this end, we consider  a point-to-point EH system that operates over a slow time-continuous fading channel with complex-valued additive white Gaussian noise (AWGN)  having unit variance.   For this system, let the square of the magnitude of the fading gain  in the $i$-th time slot  be denoted by $\gamma(i)$. Furthermore,  we assume that   the average power harvested by the EH transmitter is $\bar P_{\rm in}$. In the following,  we consider two examples with different utility functions $U(i)$.

\textit{Example 1:} For the first example, $U(i)$ is  the outage indicator for the $i$-th codeword, which is given by
\begin{eqnarray}
     U(i)=\left\{
\begin{array}{ll}
1 & \textrm{if } \log_2(1+P_{\rm out}(i)\gamma(i))< R_0\\
0 & \textrm{if } \log_2(1+P_{\rm out}(i)\gamma(i))\geq R_0 ,\\
\end{array}
\right. 
\end{eqnarray}
i.e., the value of $U(i)$ indicates whether or not an outage  occurs when the EH transmitter transmits a codeword with a fixed data rate  $R_0$.
Hence, $\bar U$ represents the fraction of codewords received  in  outage, or in other words, the outage probability of a codeword.

We would like to   minimize $\bar U$. Hence, Remark 2 applies and  the maximization in (\ref{MPR1}) has to be replaced by a minimization. Furthermore, for this example, we   impose a constraint  on $P_{\rm d}(i)$. Namely,   $P_{\rm d}(i)$ can take any value but has to be constant for all time slots. Moreover,  $\bar P_{\rm lim}\geq \bar P_{\rm in}$ is assumed. According to Theorem~\ref{theo_1}, in order to find the optimal $P_{\rm d}(i)$ which maximizes $\bar U$, we have to solve the non-EH problem in (\ref{MPR1_non-EH}) with the maximization replaced by a minimization, and with $\bar P_{\rm lim,non-EH}=\bar P_{\rm in}$ and $P_{\rm d}(i)$ being constant $\forall i$. The solution to this non-EH problem is   straightforward and given by $P_{\rm d}(i)=\bar P_{\rm in}$, $\forall i$, which, according to Theorem~\ref{theo_1}, is also the solution of the considered EH problem in (\ref{MPR1}). Hence, the output power for the $i$-th codeword in the EH system is $P_{\rm out}(i)=\min\{B(i-1),P_{\rm d}(i)\}=\min\{B(i-1),\bar P_{\rm in}\}$.

\textit{Example 2:} For the second example, let  $U(i)$ represent  the maximum information  rate of the $i$-th codeword. To underline the generality of the proposed framework, we   consider now an example that accounts for power amplifier inefficiency  \cite{5722444,kwan_energy}. In particular,  we model the "transmit" power\footnote{In this case, $P_{\rm out}(i)$ is actually the power drawn from the battery and consumed for the transmission of the $i$-th codeword.} for the $i$-th codeword as \cite{kwan_energy}
\begin{eqnarray}\label{ll_1}
    P_{\rm out}(i)=\varepsilon P_{\rm out}^T(i) +P_{\rm C},
\end{eqnarray}
where  $P_{\rm out}^T(i)$ is the actual transmit power of the $i$-th codeword and  $\varepsilon\geq 1$ is a constant which accounts for the inefficiency of the power amplifier. For example, if $\varepsilon = 5$, then $5$ Watts are consumed in
the power amplifier and have to be drawn from the battery for every $1$ Watt of power radiated in the
RF, which results in a power efficiency of $1/\varepsilon = 1/5 = 20\%$. The power
that is not radiated is dissipated as heat in the power amplifier \cite{kwan_energy}. Furthermore,  $P_{\rm C}\geq 0$  is the static circuit power
consumption of the transmitter  device electronics  such as
mixers, filters, digital-to-analog converters, and is independent of the actual transmitted power $P_{\rm out}^T(i)$. Hence,  $U(i)$ is given by 
\begin{align}
    U(i)&=\log_2\left(1+P_{\rm out}^T(i)\gamma(i)\right)\nonumber\\
&= \log_2\left(1+\frac{1}{\varepsilon}(P_{\rm out}(i)-P_{\rm C})^+\gamma(i)\right),
\end{align}
where $(x)^+=\max\{0,x\}$. Given  $U(i)$, $\bar  U$ is the maximal  average data rate  that the EH transmitter can transmit to the receiver. For this example, we do not impose any additional constraints on $P_{\rm d}(i)$ and assume $\bar P_{\rm lim}\geq\bar P_{\rm in}$. According to Theorem~\ref{theo_1}, the optimal power allocation for this EH system can be found by solving  the non-EH optimization problem in  (\ref{MPR1_non-EH}) with $\bar P_{\rm lim,non-EH}=\bar P_{\rm in}$. To this end, we follow the  approach  in \cite{goldsmith1997capacity} (where $\varepsilon=1$ and $P_C=0$ was assumed) and obtain the   solution to the non-EH optimization problem in  (\ref{MPR1_non-EH}) as
\begin{eqnarray}\label{eq_p_1a}
    P_{\rm d}(i)= 
\left\{
\begin{array}{ll}
P_{\rm C} + \varepsilon (1/\lambda-1/\gamma(i)), &\textrm{if } \;\gamma(i)>\lambda\\
0,& \textrm{if } \; \gamma(i)\leq\lambda,
\end{array}
\right.
\end{eqnarray}
where   $\lambda$ is found as the solution to 
$E\{P_{\rm d}(i)\} = \bar P_{\rm in}$. According to Theorem~\ref{theo_1}, the optimal desired power of the non-EH system in (\ref{eq_p_1a})  is also the desired power for the  EH system.  Hence, $P_{\rm out}(i)$ is given by $P_{\rm out}(i)=\min\{B(i-1),P_{\rm d}(i)\}$, where $P_{\rm d}(i)\}$ is given in  (\ref{eq_p_1a}).

%Since for this application if $P_{\rm out}(i)< P_{\rm d}(i)$, a codeword with rate $\log_2(1+P_{\rm out}^T(i))$ cannot be decoded at the destination, we choose not to transmit a codeword when this even occurs. Hence, $P_{\rm out}(i)$ is given by 
%\begin{eqnarray}\label{eq_p_121a}
%    P_{\rm out}(i)=
%\left\{
%\begin{array}{ll}
% P_{\rm C} +  \varepsilon (1/\lambda-1/\gamma(i)) , &\textrm{if } \;\gamma(i)\geq\lambda \textrm{ AND } B(i-1)\geq P_{\rm d}(i)\\
%0,& \textrm{if } \; \gamma(i)<\lambda\textrm{ OR } B(i-1)<P_{\rm d}(i) .
%\end{array}
%\right.\nonumber
%\end{eqnarray}

Numerical results for  Examples~1 and 2 are provided in Section V.

\section{The Broadcast and Multiple-Access EH Networks}
In the following, we generalize the framework developed for the point-to-point EH channel  to  the broadcast (point-to-multipoint) and the multiple-access  (multipoint-to-point) EH networks. Thereby, we  show that the maximum  average performances of the broadcast  and  multiple-access EH networks converge to the maximum average performance  of their equivalent broadcast  and  multiple-access non-EH networks, respectively.

\subsection{The Broadcast EH Network}
Let us   assume a single EH transmitter transmitting  to $M$   receiving nodes (receivers).  In each time slot, the EH transmitter extracts power from its battery and uses it to  transmit codewords to each of the receivers. Let the transmit power of the codeword  transmitted to the $k$-th receiver in the $i$-th time slot be denoted by $P_{{\rm out},k}(i)$\footnote{If the same codeword is transmitted to more than one receiver, then, in terms of transmit power, these receivers can be merged into a single equivalent receiver.}. Without loss of generality, we assume that in each time slot the EH transmitter  extracts power from the battery in the following predefined sequential manner. In each time slot, the transmitter first extracts the power for the codeword transmitted  to the first receiver, then, from the leftover power in the battery it extracts the power for the codeword transmitted  to the second receiver, and so on until from the leftover power in its battery it extracts the power for the codeword transmitted to the $M$-th receiver. Let $P_{{\rm d},k}(i)\in\mathcal{P}$ denote the desired transmit power for the codeword to the $k$-th receiver in the $i$-th time slot. Then,  $P_{{\rm out},k}(i)$ is given by
\begin{eqnarray}\label{eq_PP_n}
    P_{{\rm out},k}(i)=\left\{\hspace{-1.5mm}
\begin{array}{ll}
P_{{\rm d},k}(i),  \textrm{ if } B(i-1)\geq \sum_{j=1}^k P_{{\rm d},j}\\
B(i-1)- \sum_{j=1}^{k-1} P_{{\rm out},j},    \textrm{ otherwise},
\end{array}
\right.
\end{eqnarray}
where $B(i)$ is  the amount of power in the  battery of the EH transmitter in the $i$-th time slot and is given by
\begin{eqnarray}\label{eq_PP_n12}
    B(i)=B(i-1)+P_{\rm in}(i)-\sum_{k=1}^M P_{{\rm out},k}(i).
\end{eqnarray}
Here, $P_{\rm in}(i)$ is the amount of power  harvested  in the $i$-the time slot.
The total transmit power and the desired total transmit power of the EH transmitter in time slot $i$, denoted by $P_{{\rm out}}(i)$ and  $P_{\rm d}(i)$, respectively, are given by
\begin{align}
    P_{{\rm out}}(i) = \sum_{k=1}^M P_{{\rm out},k}(i),\quad
    P_{{\rm d}}(i) = \sum_{k=1}^M P_{{\rm d},k}(i).
\end{align}
Hence, the average transmit power and the average desired power of the EH transmitter, denoted by $\bar P_{{\rm out}}$ and $\bar P_{{\rm d}}$, respectively, are given by 
\begin{align}
   \bar P_{{\rm out}} =& \lim_{N\to\infty}\frac{1}{N} \sum_{i=1}^N P_{{\rm out}}(i) = \lim_{N\to\infty} \frac{1}{N} \sum_{i=1}^N  \sum_{k=1}^M P_{{\rm out},k}(i)\\
   \bar P_{{\rm d}} =& \lim_{N\to\infty} \frac{1}{N} \sum_{i=1}^N P_{{\rm d}}(i) = \lim_{N\to\infty} \frac{1}{N} \sum_{i=1}^N  \sum_{k=1}^M P_{{\rm d},k}(i).
\end{align} 
Using  $P_{\rm in}(i)$, the average harvested power, $\bar P_{\rm in}$, is given  by (\ref{eq_P_in}).
Now, let $\bar P_{\rm lim}$ denote the upper limit on the average transmit power $\bar P_{\rm out}$. Then, the average transmit power  must satisfy
$\bar P_{{\rm out}} \leq \bar P_{\rm lim}$. In the following, we introduce the utility function of the broadcast network which is a multivariate version of the utility function of the point-to-point system.  

 Let $U(i)$ denote the utility function of the broadcast network in the $i$-th time slot. The utility function $U(i)$ is now associated with all $M$ codewords transmitted in the $i$-th time slot, and, similar to the point-to-point case, it measures some desired quality of the codewords transmitted in the  $i$-th time slot.  Let $U(i)$ be a function of all $M$ transmit powers $P_{{\rm out},k}(i)$, $k=1,...,M$. We formally express the dependence of the utility function on the $M$ transmit powers as $U(P_{{\rm out},1}(i),P_{{\rm out},2}(i),...,P_{{\rm out},M}(i))$. For simplicity of presentation, we write $U(i)$ instead of $U(P_{{\rm out},1}(i),P_{{\rm out},2}(i),...,P_{{\rm out},M}(i))$. We assume that  the properties of $U(i)$ as a function of an individual transmit power $P_{{\rm out},k}(i)$, $\forall k=1,...,M$, are as outlined in Definition~\ref{def_1}. Given $U(i)$, $\forall i$, the average utility function $\bar U$ can be found using (\ref{av_ut_fun}). Based on $\bar U$, we introduce  now  the power allocation problem for the broadcast EH network.

For the broadcast EH network, given a limit on the average transmit power, $\bar P_{\rm lim}$, we wish to  determine the optimal desired powers $P_{{\rm d},k}(i)$, $\forall i,k$, which produce the corresponding $P_{{\rm out},k}(i)$, $\forall i,k$,  such that $\bar P_{\rm out}\leq \bar  P_{\rm lim}$ holds and the average utility function  $\bar U$  is maximized.  We define this rigorously in the following maximization problem for $N\to\infty$
\begin{align}\label{MPR1-EH-bc}
\hspace{-2mm}
\begin{array}{cl}
& {\underset{P_{{\rm d},k}(i) \in \mathcal{P},\forall k,i }{\rm{Maximize: }}} \lim\limits_{N\to\infty}\frac{1}{N}\sum_{i=1}^N U(i)\\
&{\rm{Subject\;\; to: }}\\
 &   {\rm C1:}\;  P_{{\rm out},k}(i)=\left\{\hspace{-1.5mm}
\begin{array}{ll}
P_{{\rm d},k}(i),  \textrm{ if } B(i-1)\geq \sum_{j=1}^k P_{{\rm d},j}\\
B(i-1)- \sum_{j=1}^{k-1} P_{{\rm out},j},    \textrm{ otherwise},
\end{array}
\right. \\
&  {\rm C2:} \;  \bar P_{{\rm out}} \leq \bar P_{\rm lim} \\
 &  {\rm C3:} \;  \textrm {Optional constraints on }  P_{{\rm d},k}(i)   \\
& {\rm C4:}\; B(i)=B(i-1)+P_{\rm in}(i)-\sum_{k=1}^M P_{{\rm out},k}(i),
\end{array}
\end{align} 
where $P_{\rm in}(i)$ is known causally at the EH transmitter as explained for the point-to-point EH channel. 
  
On the other hand, for the equivalent non-EH broadcast network, the non-EH transmitter can supply any desired power, thus $P_{{\rm out},k}(i)=P_{{\rm d},k}(i)$, $\forall i,k$. Therefore, maximizing the average utility function, $\bar U$, for the equivalent non-EH broadcast   network has the following form for $N\to\infty$
\begin{eqnarray}\label{MPR1_non-EH-bc}
\hspace{-2mm}
\begin{array}{cl}
 {\underset{P_{{\rm d},k}(i)\in \mathcal{P},\forall k,i }{\rm{Maximize: }}}&\lim\limits_{N\to\infty}  \frac{1}{N}\sum_{i=1}^N U(i)\\
{\rm{Subject\;\; to: }} &  {\rm C1:}\; P_{{\rm out},k}(i)=P_{{\rm d},k}(i) \;  \\
 &  {\rm C2:} \;\bar P_{{\rm d}} \leq \bar P_{\rm lim,non-EH}\\
& {\rm C3:}\; \textrm {Optional constraints on }  P_{{\rm d},k}(i), 
\end{array}
\end{eqnarray} 
where $U(i)$, $\forall i$,  and C3 are the same as in  (\ref{MPR1-EH-bc}).
We  now present the   solution of (\ref{MPR1-EH-bc})   in the following theorem.
\begin{theorem}\label{theo_3}
The solution of the non-EH optimization problem in (\ref{MPR1_non-EH-bc}) with $\bar P_{\rm lim,non-EH}=\min\{\bar P_{\rm lim}, \bar P_{\rm in}\}$ is also the solution to the EH optimization problem in (\ref{MPR1-EH-bc}). As a result,  the maximum average performance of an EH broadcast  network, $\bar U$, is identical to the maximum average performance of its equivalent non-EH broadcast  network  with $\bar P_{\rm lim,non-EH}=\min\{\bar P_{\rm lim}, \bar P_{\rm in}\}$. 
\end{theorem} 
\begin{IEEEproof}
Please refer to Appendix~\ref{app_3}.
\end{IEEEproof} 
Next, we consider the multiple-access EH network.
\subsection{The Multiple-Access EH Network}
The multiple-access EH network is comprised of $M$ EH transmitters transmitting   to a single receiving node (receiver). Here, we impose one  practical constraint by assuming that the harvested energy from one EH transmitter cannot be transferred to another EH transmitter. In each time slot, each EH transmitter extracts power from its battery and uses it to  transmit a codeword  to the receiver. Let the transmit power of the codeword  transmitted by the $k$-th EH transmitter in the $i$-th time slot be denoted by $P_{{\rm out},k}(i)$.  Let $P_{{\rm d},k}(i)\in\mathcal{P}$ denote  the desired power that the $k$-th EH transmitter wants to extract from its battery in the $i$-th time slot. 
Then,  $P_{{\rm out},k}(i)$ and $P_{{\rm d},k}(i)$  are related by
\begin{eqnarray} \label{eq_PP_nn}
    P_{{\rm out},k}(i)=\min\{B_k(i-1),P_{{\rm d},k}(i)\},
\end{eqnarray}
where $B_k(i)$ is  the amount of power in the battery of the $k$-th EH transmitter  in the $i$-th time slot and is given by
\begin{eqnarray} \label{eq_PP_nn12}
    B_k(i)=B_k(i-1)+P_{{\rm in},k}(i)-P_{{\rm out},k}(i).
\end{eqnarray}
Here, $P_{{\rm in},k}(i)$ is the harvested power at the $k$-th EH transmitter in the $i$-th time slot.
For the $k$-th EH transmitter, the average transmit power, denoted by $\bar P_{{\rm out},k}$, the average desired power, denoted by $\bar P_{{\rm d},k}$, and the average harvested power, denoted by $\bar P_{{\rm in},k}$, are given by 
\begin{eqnarray}
   \bar P_{{\alpha},k}& =& \lim_{N\to\infty}\frac{1}{N} \sum_{i=1}^N P_{{\alpha},k}(i), \quad \alpha\in\{{\rm out, d, in }\}.
\end{eqnarray} 
Furthermore, each EH transmitter imposes an  upper limit on the average transmit power, which for the $k$-th EH transmitter is denoted by $\bar P_{{\rm lim},k}$. Thus, $\bar P_{{\rm out},k}\leq \bar P_{{\rm lim},k}$ has to hold.

Similar to the broadcast EH network,  the utility function of the multiple-access EH network in the $i$-th time slot, $U(i)$,   depends on all $P_{{\rm out},k}(i)$, $k=1,...,M$, and, as a function of any individual $P_{{\rm out},k}(i)$,   has   the properties laid out in Definition~\ref{def_1}.  

For the multiple-access EH network, given a limit on the average transmit power of each EH transmitter, $\bar P_{{\rm lim},k}$, we wish  to determine the optimal desired powers $P_{{\rm d},k}(i)$, $\forall i,k$, which produce the corresponding   $P_{{\rm out},k}(i)$, $\forall i,k$,  such that $\bar P_{{\rm out},k}\leq \bar  P_{{\rm lim},k}$,  $\forall k$, holds, and the average utility function  $\bar U$  is maximized.  This leads to the following maximization problem 
\begin{align}\label{MPR1-EH-mac}
\hspace{-2mm}
\begin{array}{cl}
& {\underset{P_{{\rm d},k}(i)\in\mathcal{P},\forall k,i }{\rm{Maximize: }}} \lim\limits_{N\to\infty} \frac{1}{N}\sum_{i=1}^N U(i)\\
&{\rm{Subject\;\; to: }} \\
&   {\rm C1:}\; P_{{\rm out},k}(i)=\min\{B_k(i-1),P_{{\rm d},k}(i)\}  \\
&  {\rm C2:} \;  \bar P_{{\rm out},k} \leq \bar P_{{\rm lim},k}  \\
 &  {\rm C3:} \;  \textrm {Optional constraints on }  P_{{\rm d},k}(i) \\
& {\rm C4:}\; B_k(i)=B_k(i-1)+P_{{\rm in},k}(i)-P_{{\rm out},k}(i),
\end{array}
\end{align} 
where $P_{{\rm in},k}(i)$ is known causally only at the $k$-th EH transmitter.
On the other hand, for the equivalent non-EH multiple-access  network, since 
$P_{{\rm out},k}(i)=P_{{\rm d},k}(i)$, $\forall k,i$,  we have the following optimization problem 
\begin{eqnarray}\label{MPR1_non-EH-mac}
\hspace{-2mm}
\begin{array}{cl}
 {\underset{P_{{\rm d},k}(i)\in\mathcal{P},\forall k,i }{\rm{Maximize: }}}&\lim\limits_{N\to\infty} \frac{1}{N}\sum_{i=1}^N U(i)\\
{\rm{Subject\;\; to: }} &  {\rm C1:}\; P_{{\rm out},k}(i)=P_{{\rm d},k}(i) \\
 &  {\rm C2:} \;\bar P_{{\rm d},k} \leq \bar P_{{\rm lim,non-EH},k} \\
& {\rm C3:}\; \textrm {Optional constraints on }  P_{{\rm d},k}(i) ,
\end{array}
\end{eqnarray} 
where $U(i)$, $\forall i$, and C3 are the same as in  (\ref{MPR1-EH-mac}).
We now characterize the solution of (\ref{MPR1-EH-mac}). 
 
\begin{theorem}\label{theo_4}
The solution of the non-EH optimization problem in (\ref{MPR1_non-EH-mac}) with $\bar P_{{\rm lim,non-EH},k}=\min\{\bar P_{{\rm lim},k}, \bar P_{{\rm in},k}\}$ is also the solution to the EH optimization problem in (\ref{MPR1-EH-mac}). As a result,  the maximum average performance of an EH multiple-access network, $\bar U$, is identical to the maximum average performance of its equivalent non-EH multiple-access network with $\bar P_{{\rm lim,non-EH},k}=\min\{\bar P_{{\rm lim},k}, \bar P_{{\rm in},k}\}$. 
\end{theorem}

\begin{IEEEproof}
Please see Appendix~\ref{app_4}.
\end{IEEEproof}
In the following, we present examples for power allocation in the    multiple-access EH network.

\subsection{Example for Power Allocation in Broadcast  EH Networks}

\textit{Example 3:}
We assume a   broadcast EH  network comprised of an EH transmitter and two   receivers, where the receivers are impaired by complex-valued  AWGN with unit variance. We assume that the channel gains from the EH transmitter to the two receivers are fixed and denoted by $\sqrt{\gamma_k}$, $k=1,2$.  Let the average power harvested by the EH transmitter  be $\bar P_{\rm in}$, where  $\bar P_{\rm lim}\geq \bar P_{\rm in}$ holds. Our goal is  to obtain the     capacity region of this EH network using the proposed framework.   

 Assuming $\gamma_1<\gamma_2$, the maximum  achievable rates for receivers 1 and 2 in time slot $i$, denoted by  $R_1(i)$ and $R_2(i)$, respectively, are given by \cite{cover2006elements} 
\begin{align} 
R_1(i)&=\log_2\left(1+\frac{P_{{\rm out},1}(i)\gamma_1}{P_{{\rm out},2}(i)\gamma_1+1} \right)\label{eq_eh1a}\\
R_2(i)&=\log_2\left(1+ P_{{\rm out},2}(i)\gamma_2 \right). \label{eq_eh1b}
\end{align}
 The rates  (\ref{eq_eh1a}) and (\ref{eq_eh1b}) are achieved in the following manner \cite{cover2006elements}.  In time slot $i$,  the EH transmitter transmits a superimposed codeword $X(i)$ comprised of two  codewords $X_1(i)$ and $X_2(i)$ as  $X(i)=X_1(i)+X_2(i)$. The codewords $X_1(i)$ and $X_2(i)$ contain $n\to\infty$ symbols, where each symbol is generated independently from a zero-mean complex-valued Gaussian distribution with variances $P_{{\rm out},1}(i)$  and $P_{{\rm out},2}(i)$, respectively. On the other hand, in time slot $i$, receivers 1 and 2  receive codewords $Y_1(i)$ and $Y_2(i)$, respectively, given by
\begin{align}
Y_1(i)&=\sqrt{\gamma_1} X_1(i)+\sqrt{\gamma_1} X_2(i) + N_1(i)\label{eq_eh2a}\\
Y_2(i)&=\sqrt{\gamma_2} X_1(i)+\sqrt{\gamma_2} X_2(i) + N_2(i)\label{eq_eh2b},
\end{align}
where $N_1(i)$ and $N_2(i)$ are the unit-variance complex-valued AWGNs at receivers 1 and 2, respectively. The decoding at the receivers is performed as follows. Receiver 1 decodes  $X_1(i)$ by considering $X_2(i)$ as interference. Since in this case the resulting channel is a complex-valued AWGN channel with SNR $P_{{\rm out},1}(i)\gamma_1/(P_{{\rm out},2}(i)\gamma_1+1)$, the decoding of $X_1(i)$ at receiver 1 is successful  if the rate of $X_1(i)$ is  smaller than or equal  to  $R_1(i)$, see \cite{cover2006elements}. Similarly, receiver 2 also decodes  $X_1(i)$ by considering $X_2(i)$ as interference.  Since, in this case, the resulting channel is a complex-valued AWGN channel with SNR $P_{{\rm out},1}(i)\gamma_2/(P_{{\rm out},2}(i)\gamma_2+1)$, the decoding of $X_1(i)$ is successful   if the rate of $X_1(i)$ is smaller than or equal  to $R_3(i)=\log_2\left(1+\frac{P_{{\rm out},1}(i)\gamma_2}{P_{{\rm out},2}(i)\gamma_2+1} \right)$. Now, since   the rate of $X_1(i)$ is $R_1(i)$, and since for $\gamma_1<\gamma_2$, $R_1(i)<R_3(i)$ holds,  receiver 2 can decode $X_1(i)$ successfully.  Once receiver 2 has decoded $X_1(i)$, it subtracts $\gamma_2 X_1(i)$ from the received codeword $Y_2(i)$ and thereby obtains a new received codeword, denoted by $Y_2'(i)$, as 
\begin{align}\label{eq_eh3}
Y_2'(i)&=Y_2 (i)- \gamma_2 X_1(i) =\gamma_2 X_2(i) + N_2(i).
\end{align}
Since (\ref{eq_eh3}) is a complex-valued  AWGN channel with SNR $P_{{\rm out},2}(i)\gamma_2$, receiver 2 can decode $X_2(i)$   successfully if the rate of $X_2(i)$ is smaller than or equal to (\ref{eq_eh1b}). 

 Now, what are the optimal values for $P_{{\rm out},1}(i)$ and $P_{{\rm out},2}(i)$, $\forall i$, which   maximize the rate region? This is investigated in the following. 

Let us define the weighted sum rate $U(i)=\alpha R_1(i)+(1-\alpha) R_2(i)$, with weight $\alpha$, $0\leq \alpha\leq 1$. Then,   
\begin{align}\label{eq_eh4}
\bar  U&=\lim_{N\to\infty}\frac{1}{N}\sum_{i=1}^N U(i) \\
&= \alpha  \lim_{N\to\infty}\frac{1}{N}\sum_{i=1}^N R_1(i) + (1-\alpha)  \lim_{N\to\infty}\frac{1}{N}\sum_{i=1}^N R_2(i) \nonumber
\end{align}
  is  the   weighted  sum  rate achieved during $N\to\infty$ time slots. By maximizing $\bar  U$ for a fixed $\alpha$, we obtain one point on the boundary of the rate region, see \cite{cover2006elements}. Then, by  varying $0\leq \alpha\leq 1$, we can obtain all points on the boundary of the rate region. Now, to maximize  $\bar  U$, we insert  (\ref{eq_eh4}) into (\ref{MPR1-EH-bc}) and thereby obtain the EH optimization problem. To solve this optimization problem, we use   Theorem~\ref{theo_3} and thereby transform an EH optimization  problem   into the non-EH optimization problem in (\ref{MPR1_non-EH-bc}) with $\bar U$ replaced by (\ref{eq_eh4}) and $\bar P_{\rm lim,non-EH}$ set to $\bar P_{\rm lim,non-EH}=\bar P_{\rm in}$.  The solution of   the resulting non-EH optimization problem is given by \cite{cover2006elements}  
\begin{align}  
    P_{{\rm d},1}(i) = \alpha \bar P_{\rm in} \textrm{ and }
    P_{{\rm d},2}(i) = (1-\alpha) \bar P_{\rm in}, \; \forall i.\label{eq_1_1_1b}
\end{align}
As a result of Theorem~\ref{theo_3},  (\ref{eq_1_1_1b}) is also the solution to the EH optimization problem in (\ref{MPR1-EH-bc}). According to this solution, $P_{{\rm out},1}(i)$ and $P_{{\rm out},2}(i)$   are given  by
\begin{align}
P_{{\rm out},1}(i)&=\left\{\hspace{-1.5mm}
\begin{array}{ll}
\alpha \bar P_{\rm in},  \textrm{ if } B(i-1)\geq  \alpha \bar P_{\rm in}\\
B(i-1),    \textrm{ otherwise},
\end{array}
\right.\label{eq_1_1_1c}\\
P_{{\rm out},2}(i)&=\left\{\hspace{-1.5mm}
\begin{array}{ll}
(1-\alpha) \bar P_{\rm in},  \textrm{ if } B(i-1)\geq \bar P_{\rm in}\\
B(i-1)-   P_{{\rm out},1}(i),    \textrm{ otherwise}.\label{eq_1_1_1d}
\end{array}
\right.
\end{align}
  To show that indeed with the power allocation in (\ref{eq_1_1_1c}) and (\ref{eq_1_1_1d})  the capacity region is achieved, we insert (\ref{eq_1_1_1c}) and (\ref{eq_1_1_1d})  into $\bar R_1=\lim_{N\to\infty}\frac{1}{N}\sum_{i=1}^N R_1(i)$ and $\bar R_2=\lim_{N\to\infty}\frac{1}{N}\sum_{i=1}^N R_2(i)$, where $\bar R_1$ and $\bar R_2$ are the rates for receivers 1 and 2, respectively, achieved during $N\to\infty$ time slots. Now, utilizing   Lemma~\ref{lemma_1}, which states that if $\bar P_{\rm in}=\bar P_{{\rm d},1}+\bar P_{{\rm d},2}$ holds, $P_{{\rm out},1}(i)=\alpha \bar P_{\rm in}$ and $P_{{\rm out},2}(i)=(1-\alpha) \bar P_{\rm in}$ will occur in almost all time slot $i$, we obtain
 $\bar R_1$ and $\bar R_2$ as 
\begin{align} 
\bar R_1&=\lim_{N\to\infty}\frac{1}{N}\sum_{i=1}^N \log_2\left(1+\frac{\alpha \bar P_{\rm in} \gamma_1}{(1-\alpha) \bar P_{\rm in}\gamma_1+1} \right) \nonumber\\
&=   \log_2\left(1+\frac{\alpha \bar P_{\rm in} \gamma_1}{(1-\alpha) \bar P_{\rm in}\gamma_1+1} \right)\label{eq_eh_5a}\\
\bar R_2&=\lim_{N\to\infty}\frac{1}{N}\sum_{i=1}^N \log_2\left(1+ (1-\alpha) \bar P_{\rm in} \gamma_2 \right)\nonumber\\
&= \log_2\left(1+ (1-\alpha) \bar P_{\rm in} \gamma_2 \right), \label{eq_eh_5b}
\end{align}
which is identical to the points on the boundary of the capacity region of the complex-valued AWGN non-EH broadcast network for  average prower constraint $\bar P_{\rm in}$, see \cite{cover2006elements}.  Now, since an  EH  broadcast network cannot have a better performance than its equivalent non-EH broadcast  network, it follows that (\ref{eq_eh_5a}) and (\ref{eq_eh_5b})   indeed define the capacity region of the complex-valued AWGN  EH broadcast network.

Numerical results for this example are provided in Section~\ref{sec-num}.

\section{The General EH Network} 
In this section, we extend the framework  developed in the previous sections further and derive the asymptotically optimal power allocation for a general EH network.
\subsection{Asymptotically Optimal Power Allocation}
We  consider a network comprised of $M$ EH transmitters.    For the $k$-th EH transmitter, $k=1,...,M$, in the $i$-th time slot, we denote the harvested power, the actual transmit power, the desired transmit power, and the amount of stored power in the battery  by $P_{{\rm in},k} (i)$, $P_{{\rm out},k}(i)$, $P_{{\rm d},k}(i)$, and $B_k(i)$,  respectively. Each EH transmitter uses the harvested power to transmit to its designated receiving nodes. We collect the indices of the receiving nodes of the $k$-th EH transmitter   in  set $\mathcal{R}_k$.   Then, in the $i$-th time slot, we decompose the transmit power of the $k$-th EH node, $P_{{\rm out},k}(i)$, into $|\mathcal{R}_k|$ transmit powers as 
\begin{eqnarray}\label{e_1}
    P_{{\rm out},k}(i)=\sum_{j\in \mathcal{R}_k}  P_{{\rm out},k\to j}(i),
\end{eqnarray}
where $P_{{\rm out},k\to j}(i)$ is the transmit power of the codeword sent from the $k$-th EH transmitter to its $j$-th receiver\footnote{If the same codeword is transmitted to more than one receiver, then, in terms of transmit power, these receivers can be merged into a single equivalent receiver.} in the $i$-th time slot. Similarly,  in the $i$-th time slot, we decompose the desired transmit power of the $k$-th EH transmitter, $P_{{\rm d},k}(i)$, into $|\mathcal{R}_k|$ desired transmit powers as
\begin{align}\label{e_2}
    P_{{\rm d},k}(i)=\sum_{j\in \mathcal{R}_k}  P_{{\rm d},k\to j}(i),
\end{align}
where $P_{{\rm d},k\to j}(i)\in\mathcal{P}$ is the desired transmit power for the codeword sent from the $k$-th EH transmitter to its $j$-th receiver  in the $i$-th time slot. Using (\ref{e_1}) and (\ref{e_2}), we write the average transmit power and the average desired transmit power of the $k$-th EH transmitter as 
\begin{align}\label{e_3}
  \bar P_{{\rm out},k}& =\lim_{N\to\infty}\frac{1}{N}\sum_{i=1}^N P_{{\rm out},k}(i)\nonumber\\
&=\lim_{N\to\infty}\frac{1}{N} \sum_{i=1}^N  \sum_{j\in \mathcal{R}_k}  P_{{\rm out},k\to j}(i)
\end{align}
and
\begin{align} 
 \bar P_{{\rm d},k}=\lim_{N\to\infty}\frac{1}{N} \sum_{i=1}^N P_{{\rm d},k}(i)=\lim_{N\to\infty}\frac{1}{N} \sum_{i=1}^N  \sum_{j\in \mathcal{R}_k}  P_{{\rm d},k\to j}(i), 
\end{align}
respectively. The average harvested power of the $k$-th EH transmitter is given by 
\begin{eqnarray}\label{e_4}
 \bar P_{{\rm in},k}=\lim_{N\to\infty}\frac{1}{N} \sum_{i=1}^N P_{{\rm in},k}(i) =E\{P_{{\rm in},k}(i) \}.
\end{eqnarray}
Let $\bar P_{{\rm lim},k}$ denote the upper limit on the average transmit power of the $k$-th EH transmitter. Then, $\bar P_{{\rm out},k}\leq \bar P_{{\rm lim},k}$ has to be satisfied.  

Similar  to our model for  the broadcast EH network in Section III, and without loss of generality,  we assume that in each time slot the $k$-th EH transmitter extracts   power from its battery in the following predefined sequential manner. For each time slot, the $k$-th EH transmitter first extracts the power for transmission of the codeword intended for the first receiving node in $\mathcal{R}_k$, then, from the leftover power in the battery, the $k$-th EH transmitter extracts the power for transmission of the codeword intended for the second receiving node in $\mathcal{R}_k$, and so on until,  from the leftover power in its battery, it extracts power for   transmission of the codeword intended for the $|\mathcal{R}_k|$-th receiving node in $\mathcal{R}_k$. Then,  $P_{{\rm out},k\to j}(i)$ is given by
\begin{align}\label{eq_PP_n-G}
    P_{{\rm out},k\to j}(i)=\left\{\hspace{-1.5mm}
\begin{array}{ll}
P_{{\rm d},k\to j}(i),  \textrm{ if } B_k(i-1)\geq \sum_{l\in \mathcal{R}_k, l\leq j}  P_{{\rm d},k\to l}\\
B_k(i-1)- \sum_{l\in \mathcal{R}_k, l<j} P_{{\rm out},k\to l},    \textrm{ otherwise},
\end{array}
\right.
\end{align}
where $B_k(i)$ is the amount of power stored in the battery of the $k$-th EH transmitter  at the $i$-th time slot  and is  given by
\begin{eqnarray}\label{eq_PP_n12-G}
    B_k(i)=B_k(i-1)+P_{{\rm in},k}(i)-\sum_{j\in \mathcal{R}_k } P_{{\rm out},k
\to j}(i).
\end{eqnarray}

In a communication network comprised of multiple nodes,  codewords may arrive at the intended receiver  with a certain delay from the moment of  transmission if multiple hops are involved. For example, a codeword originating from transmitter $k$ in time slot  $l$ may pass through several hops before arriving at the intended receiver in time slot $i$, where $i\geq l$. In order to model this delay, in the following, for the $i$-th time slot, we develop a generalized  utility function, $U(i)$, which may be a function of the transmit powers in time slots prior to $i$.

Let $U(i)$ denote the utility function in the $i$-th time slot. In order to  generalize the utility function, $U(i)$,   we introduce a delay $\Delta_{k\to j}$ assigned to the pair of the $k$-th EH transmitting node and the $j$-th  receiving node. The delay $\Delta_{k\to j}$   is a constant integer number satisfying $0 \leq\Delta_{k\to j}\leq N$, $\forall i$, i.e., the delay between nodes $k$ and $j$ does not change with time\footnote{We note that a similar mathematical framework could be developed for time-varying delays $\Delta_{k\to j}$. However, this would make the presentation much more involved. Hence, for simplicity, we assume $\Delta_{k\to j}$ to be constant $\forall i$.}. Now, we define the utility function for the general network, $U(i)$,  to be a function of    $P_{{\rm out},k\to j}(i-\Delta_{k\to j})$, $\forall k,j$. Using $\mathcal{R}_k$, defined previously,  we construct the following vector of powers
\begin{align}
    {\bf  P}_{{\rm out},k}(i)=\big[&P_{{\rm out},k\to 1}(i-\Delta_{k\to 1}), P_{{\rm out},k\to 2}(i-\Delta_{k\to 2}),\nonumber\\
& ...., P_{{\rm out},k\to |\mathcal{R}_k|}(i-\Delta_{k\to |\mathcal{R}_k|})\big].
\end{align}
Using  vectors ${\bf  P}_{{\rm out},k}(i)$, for $k=1,...,M$, we express the dependence of  $U(i)$ on $P_{{\rm out},k\to j}(i-\Delta_{k\to j})$, $\forall k,j$, as\footnote{For simplicity of presentation, we have assumed that $U(i)$   depends only on the transmit power of node $k$ to node $j$ in one time instant, $\forall k,j$. However, the proposed framework can also be extend to the case when $U(i)$ depends on the transmit powers  of node $k$ to node $j$ in multiple time slots, as would be the case if, for example, an automatic repeat request (ARQ) protocol was employed. To this end, defining a corresponding equivalent  non-EH network is required.}  
\begin{eqnarray}\label{eq_37}
    U(i)=U\big( {\bf  P}_{{\rm out},1}(i),  {\bf  P}_{{\rm out},2}(i), ...,  {\bf  P}_{{\rm out},M}(i)
\big).
\end{eqnarray}
We note that   $U(i)$ as a function of any individual $P_{{\rm out},k\to j}(i-\Delta_{k\to j})$ has the  properties outlined for the point-to-point case in Definition~\ref{def_1}. Adopting $U(i)$ in (\ref{eq_37}), the average utility function $\bar U$ is given by  (\ref{av_ut_fun}).

For the general EH network, given a limit on the average transmit power of each EH transmitter, $\bar P_{{\rm lim},k}$, our objective is to determine the optimal desired powers $P_{{\rm d},k\to j}(i-\Delta_{k\to j})$, $\forall k,j,i$, which produce the corresponding transmit powers $P_{{\rm out},k\to j}(i-\Delta_{k\to j})$,  such that $\bar P_{{\rm out},k}\leq \bar  P_{{\rm lim},k}$ hold $\forall k$ and the average utility function  $\bar U$  is maximized.  We define this rigorously in the following maximization problem 
\begin{align}\label{MPR1-EH-G}
\hspace{-6mm}
\begin{array}{cl}
& {\underset{P_{{\rm d},k\to j}(i)\in\mathcal{P},\forall k,j,i }{\rm{Maximize: }}} \lim\limits_{N\to\infty} \frac{1}{N}\sum_{i=1}^N U(i)\\
&{\rm{Subject\;\; to: }}  \\
&  {\rm C1:}\; P_{{\rm out},k\to j}(i)=\left\{\hspace{-1.5mm}
\begin{array}{ll}
P_{{\rm d},k\to j}(i),  \textrm{ if } B_k(i-1)\geq \hspace{-2mm}\sum\limits_{l\in \mathcal{R}_k, l\leq j} \hspace{-4mm} P_{{\rm d},k\to l}\\
B_k(i-1)- \hspace{-3mm}\sum\limits_{l\in \mathcal{R}_k, l<j}\hspace{-4mm}  P_{{\rm out},k\to l},    \textrm{ otherwise},
\end{array}
\right.  \\
&  {\rm C2:} \;  \bar P_{{\rm out},k} \leq \bar P_{{\rm lim},k}\  \\
 &  {\rm C3:} \;  \textrm {Optional constraints on }  P_{{\rm d},k\to j}(i)   \\
& {\rm C4:}\; B_k(i)=B_k(i-1)+P_{{\rm in},k}(i)-\sum_{j\in \mathcal{R}_k } P_{{\rm out},k
\to j}(i),
\end{array}
\end{align} 
where $P_{{\rm in},k}(i)$ is known causally at the $k$-th EH transmitter   only. 
On the other hand, since for the equivalent non-EH system  $P_{{\rm out},k\to j}(i)=P_{{\rm d},k\to j}(i)$, $\forall i,k,j$,  the optimization problem  is given by
\begin{eqnarray}\label{MPR1_non-EH-G}
\hspace{-2mm}
\begin{array}{cl}
 {\underset{P_{{\rm d},k\to j}(i)\in\mathcal{P},\forall k,j,i }{\rm{Maximize: }}}&\lim\limits_{N\to\infty} \frac{1}{N}\sum_{i=1}^N U(i)\\
{\rm{Subject\;\; to: }} &  {\rm C1:}\; P_{{\rm out},k\to j}(i)=P_{{\rm d},k\to j}(i) \\
 &  {\rm C2:} \;\bar P_{{\rm d},k} \leq \bar P_{{\rm lim,non-EH},k}\\
& {\rm C3:}\; \textrm {Optional constraints on }  P_{{\rm d},k\to j}(i) ,
\end{array}
\hspace{-5mm}
\end{eqnarray} 
where $U(i)$, $\forall i$, and  C3 are the same as in  (\ref{MPR1-EH-G}). We are now ready to provide the framework for solving (\ref{MPR1-EH-G}). 
\begin{theorem}\label{theo_5}
The solution of the non-EH optimization problem in (\ref{MPR1_non-EH-G}) with $\bar P_{{\rm lim,non-EH},k}=\min\{\bar P_{{\rm lim},k}, \bar P_{{\rm in},k}\}$ is also the solution to the EH optimization problem in (\ref{MPR1-EH-G}). As a result,  the maximum average performance of a general EH network, $\bar U$, is identical to the maximum average performance of its equivalent general  non-EH network  with $\bar P_{{\rm lim,non-EH},k}=\min\{\bar P_{{\rm lim},k}, \bar P_{{\rm in},k}\}$. 
\end{theorem}

\begin{IEEEproof}
Please see Appendix~\ref{app_5}.
\end{IEEEproof}

\subsection{Example for General EH Network}
In the following, we present an example for online power allocation in  a general  EH network.

\textit{Example 4:}
In this example, we consider a multi-hop amplify-and-forward (AF) relay network comprised of one EH transmitter source, $M-2$ AF half-duplex EH relays, and a receiver. For simplicity, we assume that $M$ is an odd number.  We numerate the nodes with numbers such that the source node is node 1, the destination  node is node $M$, and the $M-2$ relays are indexed from 2 to $M-1$.
The network operates in slow time-continuous fading and the complex-valued AWGN at each receiver has unit variance. In the $i$-th time slot,   the square of the fading gain of the channel from the $k$-th EH node to the $(k+1)$-th EH node is denoted by $\gamma_{k\to k+1}(i)$ where $k=1,...,M-1$. We assume that all nodes have full CSI. The transmission from the source via the relays to the destination is carried out in the following manner. In odd (even) time slots, the $k$-th node, where $k$ is an odd (even) number, transmits to the $(k+1)$-th node. When a relay transmits in time slot $i$, it transmits a scaled version of the codeword received in time slot $i-1$. Let $U(i)$ represent the data rate  received at the destination at time slot $i$. Then, $U(i)$ is given by
\begin{align}\label{u_1_1a}
    U(i)=\left\{\hspace{-2mm}
\begin{array}{ll}
\log_2(1+\textrm{SNR}_{\rm eq}(i)) & \textrm{if } i\geq M-1 \textrm{ and } i \textrm{ is even}\\
0  & \textrm{otherwise,} 
\end{array}
\right.
\end{align}
where the equivalent SNR at the destination, $\textrm{SNR}_{\rm eq}(i)$, is given by \cite[Eq. (17)]{1200185}
\begin{align}
 &   \textrm{SNR}_{\rm eq}(i)=\Bigg( \hspace{1mm}\prod_{m=1}^{M-1}\bigg(1+\nonumber\\
&\frac{1}{P_{{\rm out},m\to m+1}(i\hspace{-1mm}-\hspace{-1mm}M\hspace{-1mm} +1\hspace{-1mm}+\hspace{-1mm}m)\gamma_{m\to m+1}(i\hspace{-1mm}-\hspace{-1mm}M\hspace{-1mm}+\hspace{-1mm}1\hspace{-1mm}+\hspace{-1mm}m)}\bigg)\hspace{-1mm}-1\Bigg)^{-1}\hspace{-3mm}.
\end{align}
Hence, $\bar U$ is the average data rate. Let $\bar P_{{\rm in},k}$ be the average harvested power of the $k$-th EH node, and let $\bar P_{{\rm lim},k}\geq P_{{\rm in},k}$, for $k=1,...,(M-1)/2$ and $\bar P_{{\rm lim},k}< P_{{\rm in},k}$, for $k=(M-1)/2+1,...,M-1$.  Furthermore, we impose the constraints that $P_{{\rm d},k\to k+1}(i)$ is either zero or assumes a constant value identical $\forall i$ for which it is not zero.  Then, we use Theorem~\ref{theo_5} to solve (\ref{MPR1-EH-G}). Thereby,  we solve the non-EH optimization problem  (\ref{MPR1_non-EH-G}) by setting $\bar P_{{\rm lim,non-EH},k}=\bar P_{{\rm in},k}$, for $k=1,...,(M-1)/2$ and $\bar P_{{\rm lim,non-EH},k}=\bar P_{{\rm lim},k}$, for $(M-1)/2+1,...,M-1$ and insert the constraint on $P_{{\rm d},k\to k+1}(i)$. The solution to this non-EH optimization problem is then straightforward and given as follows. For $k=1,...,(M-1)/2$, $ P_{{\rm d},k\to k+1}(i)$ is given by
\begin{align}\label{nov_1}
  P_{{\rm d},k\to k+1}(i)=\left\{
\begin{array}{cl}
2\bar P_{{\rm in},k} & \textrm{for  odd } i \textrm{ and  odd }k \\
2\bar P_{{\rm in},k} & \textrm{for  even } i\textrm{ and  even }k\\
0 & \textrm{for  even } i \textrm{ and  odd } k\\
0 & \textrm{for  odd } i \textrm{ and  even } k
\end{array}
\right. 
\end{align}
and for $k=(M-1)/2+1,...,M-1$,  $P_{{\rm d},k\to k+1}(i)$ is given by
\begin{align}\label{nov_2}
  P_{{\rm d},k\to k+1}(i)=\left\{
\begin{array}{cl}
2\bar P_{{\rm lim},k} & \textrm{for  odd } i \textrm{ and  odd }k \\
2\bar P_{{\rm lim},k} & \textrm{for  even } i\textrm{ and  even }k\\
0 & \textrm{for  even } i \textrm{ and  odd } k\\
0 & \textrm{for  odd } i \textrm{ and  even } k.
\end{array}
\right.  
\end{align}
According to  Theorem~\ref{theo_5}, the desired transmit powers in (\ref{nov_1}) and (\ref{nov_2}) are also the solution to the EH optimization problem  (\ref{MPR1-EH-G}). 
Hence, the transmit powers of the $k$-th EH node  in the $i$-th time slot are given by  $P_{{\rm out},k\to k+1}(i)=\min\{B_k(i-1), P_{{\rm d},k\to k+1}(i)\}$,  where $P_{{\rm d},k\to k+1}(i)$ for $k=1,...,(M-1)/2$ and $k=(M-1)/2+1,...,M-1$ are given by (\ref{nov_1}) and (\ref{nov_2}), respectively.

Numerical results for this example are provided in the following section.

\section{ Numerical Examples}\label{sec-num}

In the following, we discuss the applicability of the developed framework, and,  for the examples introduced in Sections II - IV, we provide numerical results for different numbers of time slots $N$  and/or battery capacities $B_{\rm \max}$.
 For all of the examples, we assume that the channels are Rayleigh fading with unit variance.  Furthermore, we assume that  the  power harvested by the EH transmitters  is an exponentially distributed random variable with mean $\bar P_{\rm in}$.    All figures are obtained via Monte-Carlo simulation and in all figures $\bar P_{\rm in}$ in dB is expressed with respect to the unit variance AWGN.

\subsection{Applicability of the Developed Framework}
In practical EH systems each EH node transmits in a finite number of time slots $N$ and has a finite battery capacity $B_{\rm max}$. In this case, for the proposed solution to be applicable,   $N$ has to be sufficiently large. The numerical examples in the following subsections  will illustrate what "sufficiently large" means in this context. In terms of $B_{\rm max}$, the proposed solution is applicable in two cases. The first case is  when the maximum capacity of the battery,  $B_{\rm max}$,   is much larger than the average harvested power $\bar P_{\rm in}$, i.e., when $B_{\rm max}\gg \bar P_{\rm in}$ holds. This is intuitive, since  if  $B_{\rm max}\gg \bar P_{\rm in}$  then the battery is practically never   fully filled. Hence, the finiteness of $B_{\rm max}$ has no effect.  The second case is when the  maximum capacity of the battery   is much larger than the upper limit on the average transmit power, i.e., when $B_{\rm max}\gg \bar P_{\rm lim}$ holds. In this case, either $\bar P_{\rm in}\leq \bar P_{\rm lim}$ or $\bar P_{\rm in}>\bar  P_{\rm lim}$ holds. If  $\bar P_{\rm in}>\bar P_{\rm lim}$ then the battery is almost always completely full any desired power can be accounted for, hence, the framework holds. Whereas, if $\bar P_{\rm in}\leq \bar P_{\rm lim}$ then    $\bar P_{\rm in}\leq \bar P_{\rm lim}\ll B_{\rm max}$ and the battery is practically  never fully filled, hence, the finiteness of $B_{\rm max}$ has no effect. 
For example, in today's mobile phones the maximum capacity of the battery is much larger than the average transmit power of a codeword. This means that $B_{\rm max}$  is much larger than the  upper limit on the average transmit power  $\bar P_{\rm lim}$. Hence, this corresponds to the second case and independent of how large the  average harvested power  is, the proposed solution is applicable in terms of $B_{\rm max}$.

Furthermore, the results derived in this paper for an infinite battery capacity constitute performance upper bounds for the case when the battery capacity is finite. Such performance  upper bounds are very useful. In particular, in practice, usually  heuristic online power allocation solutions are adopted due to practical constraints such as low  complexity and/or lack of  CSI. However, in order to evaluate how good   the proposed  heuristic  solutions are, a benchmark performance   is needed for comparison. The proposed asymptotically optimal power allocation scheme can serve as such a benchmark.

\subsection{Numerical Results for Example 1} 

\begin{figure}
\includegraphics[width=3.75in]{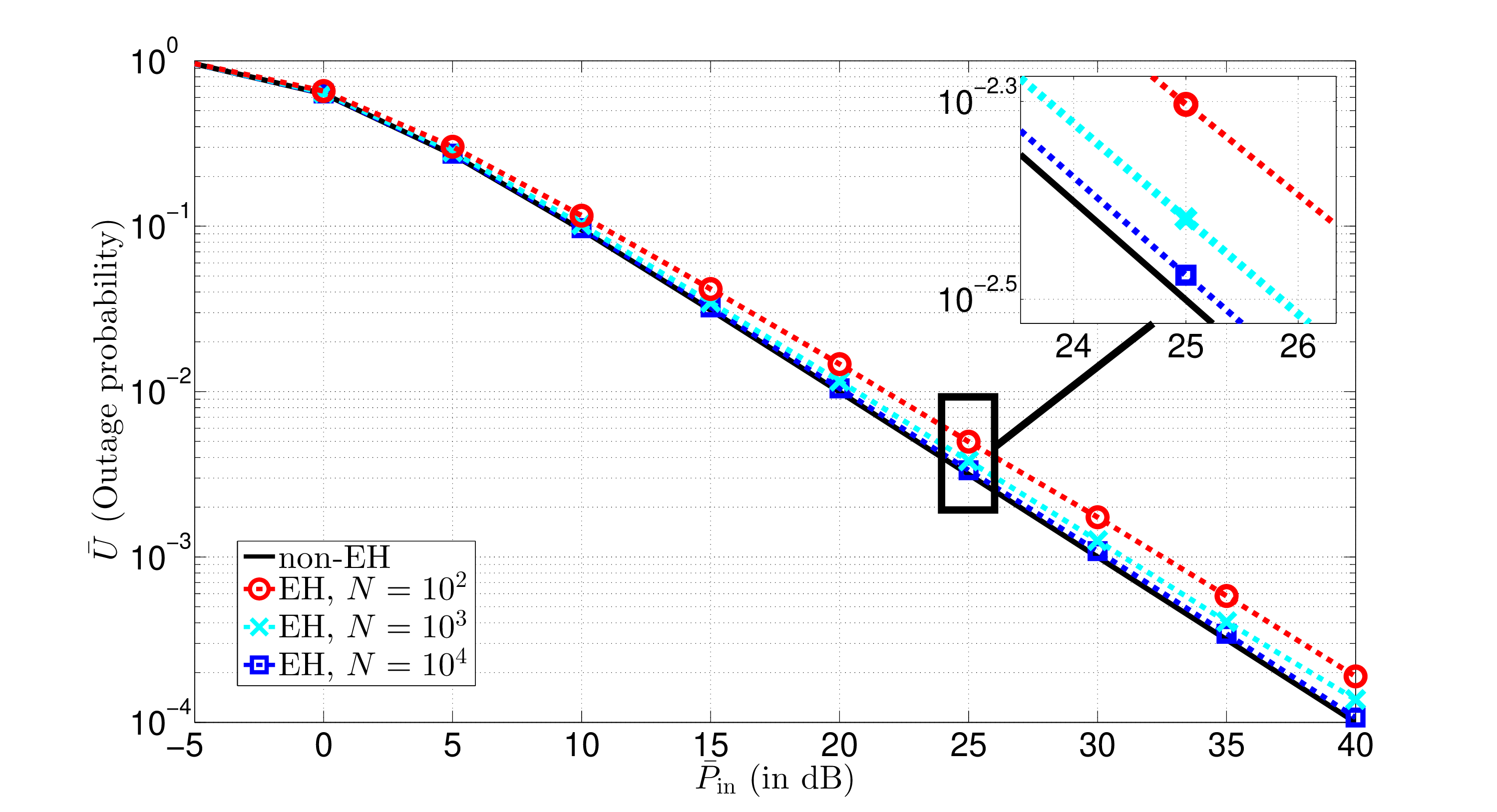}\centering
\caption{Outage probability for the point-to-point EH and the equivalent non-EH systems for different $N$ and $B_{\rm max}=200 \bar P_{\rm in}$.}\label{fig_3}
\end{figure}

In Fig.~\ref{fig_3}, we plot the  outage probability, $\bar U$, of the point-to-point  EH  system, discussed in Example 1 in Section II, for  $B_{\rm max}=200 \bar P_{\rm in}$ and different $N$,  and compare  it to the outage probability of the non-EH system with an average transmit power $\bar P_{\rm in}$ and infinite $N$.  From Fig.~\ref{fig_3}, we  observe that even for $N=10^2$, the loss in outage probability performance of the  proposed online solution  is relatively small (less than $3$ dB in the entire considered range of $\bar P_{\rm in}$) compared to the performance of the non-EH system. This performance loss becomes almost negligible for $N=10^4$. Hence, for the point-to-point EH system  the proposed online solution, although sub-optimal for finite $N$, yields a  high performance at low complexity even for small $N$.

\subsection{Numerical Results for Example 2}

\begin{figure}
\includegraphics[width=3.75in]{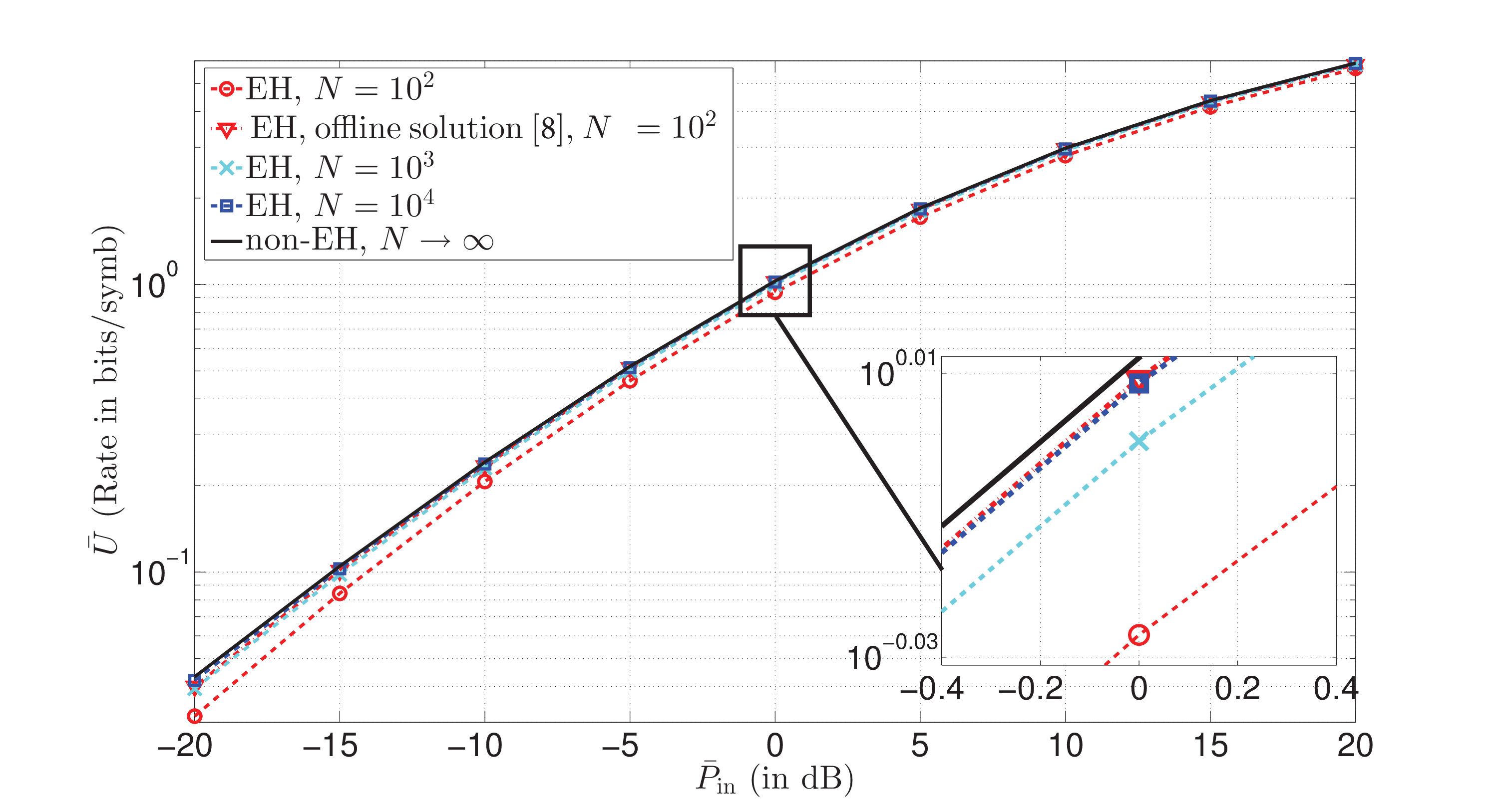}\centering
\caption{Average rates for the point-to-point EH and the equivalent non-EH systems for different $N$ and $B_{\rm max}=200 \bar P_{\rm in}$.}\label{fig_1}
\end{figure}

In   Fig.~\ref{fig_1}, we show the average data rate, $\bar U$, of the point-to-point EH system  discussed in Example 2 in Section II, for  different   $N$ and $B_{\rm max}=200 \bar P_{\rm in}$,  and compare  it to the average data rate   of the equivalent non-EH system with   average transmit power $\bar P_{\rm in}$ and infinite\footnote{Note that also the non-EH system achieves a lower  average data rate  for finite $N$   than  for $N\to\infty$, i.e., even for the non-EH system $N\to\infty$ has to be assumed for the power allocation in (\ref{eq_p_1a}) to be  optimal.} $N$. An ideal transmitter with $\varepsilon=1$ and $P_{\rm C}=0$ is assumed. 
 Fig.~2 shows that even for finite $N$ and finite $B_{\rm max}$, the loss in rate of the point-to-point    EH  system is small compared to the  non-EH system.
For $N=10^2$, as a performance upper bound,  we also show the average rate obtained with the offline solution from \cite{ho2012optimal}. We note that the offline solution, however, needs noncausal knowledge of the harvested powers and the fading gains in all $N=10^2$ time slots. In contrast, our proposed solution, although non-optimal for finite $N$, is an online solution and requires only causal knowledge of the  fading gain in the current time slot, the average harvested power, and the average fading gain. The optimal online solution for $N=10^2$  would result  in a curve which is  between the curve obtained with the optimal offline solution from \cite{ho2012optimal} and the  curve obtained with our proposed solution. However,  the optimal online solution for finite $N$ is based on dynamic  programing and its exponentially increasing computational complexity becomes prohibitive for $N=10^2$.

\begin{figure}
\includegraphics[width=3.75in]{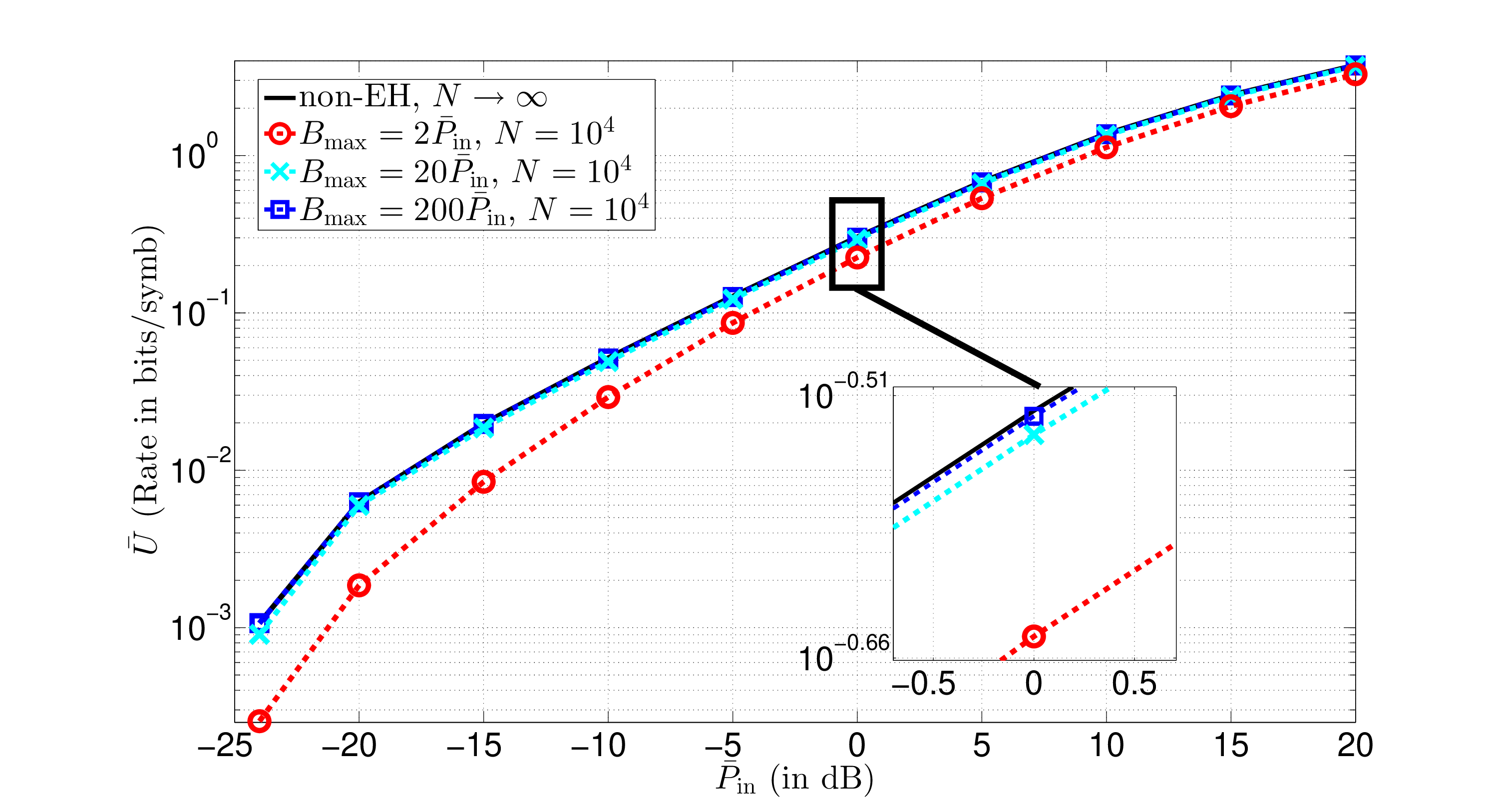}\centering
\caption{Average rates for the point-to-point EH system and the equivalent non-EH systems for different battery capacity $B_{\rm max}$, and   $N=10^4$. A realistic transmitter model is adopted with power efficiency $\varepsilon=20$\% and constant circuit power consumption $P_{\rm C}=-25$ dB.}\label{fig_2}
\end{figure}

In Fig.~\ref{fig_2}, we plot the average data rate, $\bar U$, of a    point-to-point  EH system for $\varepsilon=5$ and $P_{\rm C}=-25 \;{\rm dB}$. We plot $\bar U$ for fixed $N=10^4$ and different  $B_{\rm max}$,  and compare  it to the average data rate   of the   non-EH system with an average transmit power $\bar P_{\rm in}$ and infinite $N$. The figure shows that, for the adopted $N$, even with $B_{\rm max}=20 \bar P_{\rm in}$, the loss in rate of the  point-to-point  EH system is small compared to the non-EH system. We note that in the figure, all rates are zero for $P_{\rm in}\leq -25\;{\rm dB}$ as $P_{\rm C}=-25\;{\rm dB}$.

\subsection{Numerical Results for Example 3}
\begin{figure}
\includegraphics[width=3.5in]{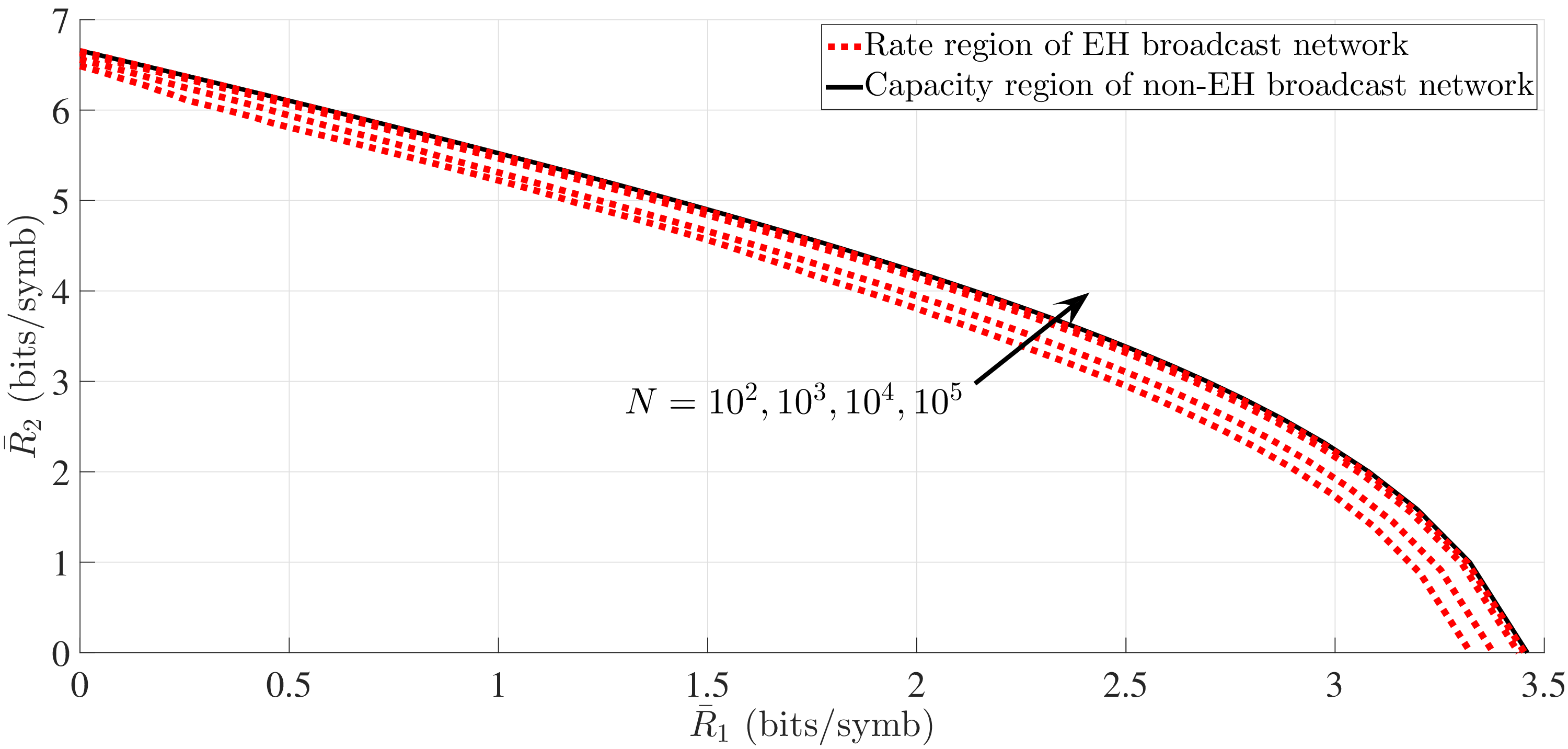}\centering
\caption{Comparison between the rate region of the broadcast EH  network for different $N$, $\gamma_1=1$, $\gamma_2=10$, and $\bar P_{\rm in}=10$, and the capacity region of the non-EH broadcast network with an average transmit power $\bar P_{\rm in}=10$.}\label{fig_5}
\end{figure}

In Fig.~\ref{fig_5}, we plot the rate region  of the broadcast  EH network with two receivers, cf. Example 3 in Section III,   for  different $N$,  and compare  it to the capacity region of the non-EH broadcast network with an average transmit power $\bar P_{\rm in}$. For this example, we set $\gamma_1=1$, $\gamma_2=10$, and $\bar P_{\rm in}=10$, cf. Example 3 in Section III.
The figure shows that  the difference between the rate region  of the broadcast EH network with $N=10^2$ and the capacity region of the non-EH broadcast network is relatively small  and becomes almost negligible for $N=10^5$.  Hence, indeed as $N\to\infty$, the  rate region  of the broadcast EH network becomes the capacity region.

\subsection{Numerical Results for Example 4}
\begin{figure}
\includegraphics[width=3.75in]{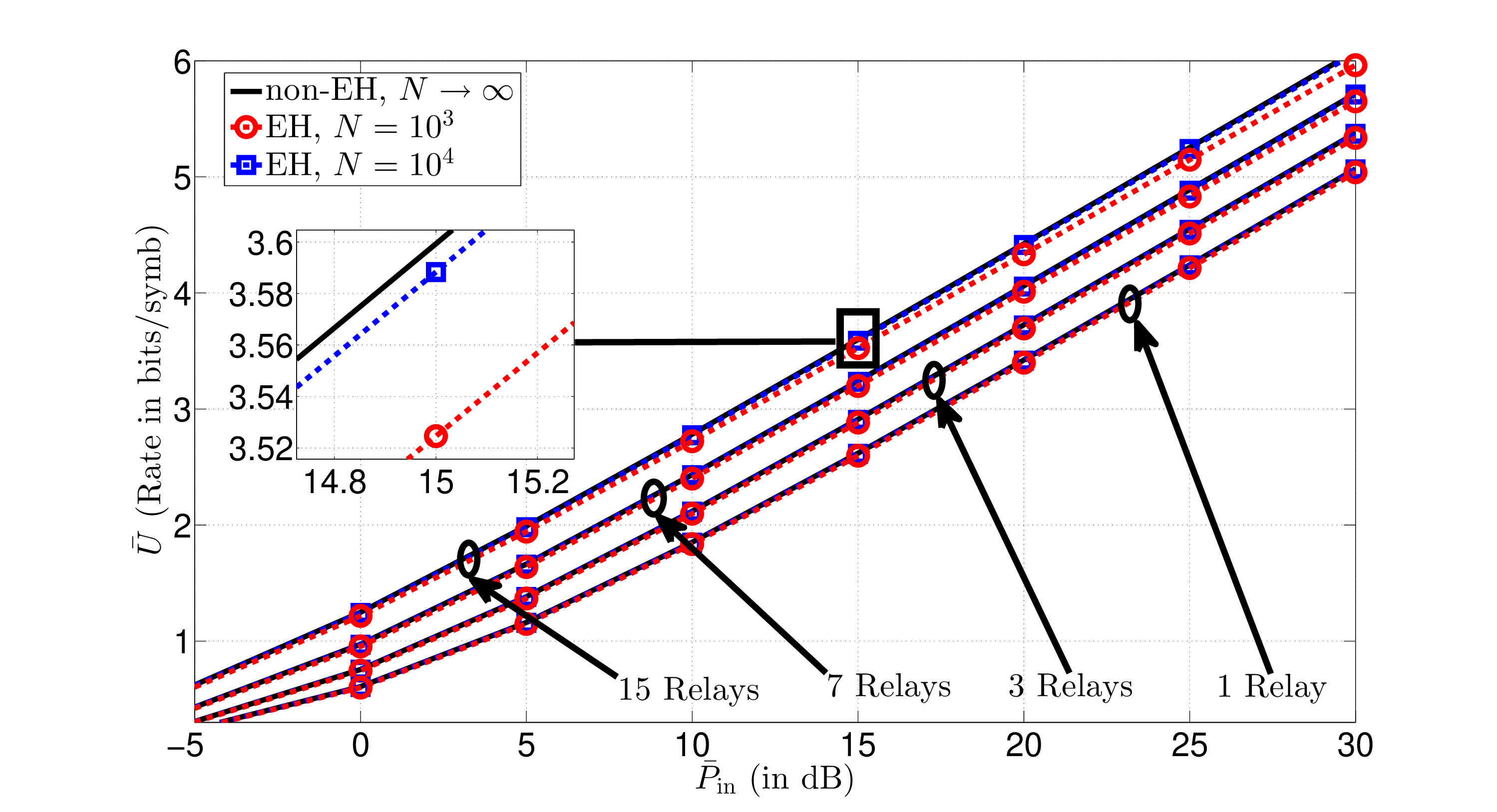}\centering
\caption{Average rate for multihop relay EH network and equivalent  multihop non-EH network for different $N$, $B_{\rm max}=200 \bar P_{\rm in}$, and different numbers of relays.}\label{fig_7}
\end{figure}

In Fig.~\ref{fig_7}, we plot the average data rate, $\bar U$, of the multihop relay EH network, considered in Example 4 in Section IV,   for  $B_{\rm max}=200 \bar P_{\rm in}$, different $N$, and different numbers of relays, and compare  it to the average data rate  of the equivalent non-EH relay  network with  infinite $N$. We assume that $\bar P_{{\rm lim},k}=\bar P_{\rm in}/2$,  for $ k=(M-1)/2+1,....,M-1$. Furthermore, we assume that the distance between  source and  destination is fixed, and that the relays are equidistantly spaced on the line between the source and destination. Hence, if we assume a path loss model with a path loss exponent equal to two, and assume that the fading gains squared have a unit mean, then, for $M-1$ relays between source and destination, the fading gains squared have a mean $M^2$. Note that each relay has an average harvested power of  $\bar P_{\rm in}$. The figure shows that, for fixed $N$, as the number of relays increases, the performance loss of  the proposed  power allocation solution also increases compared to the performance of the non-EH network. However, as $N$ increases this performance loss becomes negligible as highlighted in  Fig.~\ref{fig_7}   for $N=10^4$ and $15$ relays. Hence, even if the utility function $U(i)$ has a relatively complicated form, as is the case in this example, the performance of the proposed online solution for the general EH network is  almost identical to the performance of the equivalent non-EH network  even for  moderate $N$.

\section{Conclusion}
We have shown that the maximum average performance of an EH network, utilizing  optimal online power allocation, converges to the maximum average performance of an equivalent non-EH network with appropriately chosen average transmit power  when the number of transmit time slots, $N$, and the battery capacities at each EH node, $B_{\max}$, satisfy $N\to\infty$ and $B_{\max}\to\infty$. We have derived the asymptotically optimal online power allocation which for a general EH network optimizes a general utility function   for $N\to\infty$ and $B_{\max}\to\infty$. The considered family of utility functions is  general enough to include the most important performance measures in communication theory such as the ergodic data rate, outage probability, average bit error probability, and average signal-to-noise ratio.  The optimal online power allocation solution is obtained by solving  the power allocation problem of an equivalent non-EH network with nodes having infinite energy  available for the transmission of their codewords.  Interestingly, the optimal solution only requires knowledge of the average harvested energy but not of the amount  of harvested energy in  past, present,  or future time slots.  Although  asymptotic in nature, the proposed solution  is applicable to EH systems transmitting in a large but finite number of time slots and having nodes with battery capacities  much larger than the  average harvested power  and/or the maximum   average transmit power.

\begin{appendix}

\subsection{Proof of Lemma~\ref{lemma_1}}\label{app_1}

We divide the proof in two parts. In Parts I and II, we consider the cases  when $\bar P_{\rm d}$ is set to $\bar P_{\rm d}=\bar P_{\rm lim}<\bar P_{\rm in}$ and $\bar P_{\rm d}=\bar P_{\rm in}\leq\bar P_{\rm lim}$, respectively.

\subsubsection{Part I   ($\bar P_{\rm d}=\bar P_{\rm lim}<\bar P_{\rm in}$)} 
 
Taking the summation $\lim\limits_{N
\to\infty}\frac{1}{N}\sum_{i=1}^N(\cdot)$ of both sides of  (\ref{eq_q}), we obtain
\begin{align}\label{eq_nz_q_1}
    &\lim\limits_{N
\to\infty}\frac{1}{N}\sum_{i=1}^N B(i) =\lim\limits_{N
\to\infty}\frac{1}{N}\sum_{i=1}^N B(i-1) \nonumber\\
& +\lim\limits_{N
\to\infty}\frac{1}{N}\sum_{i=1}^N  P_{\rm in}(i)- \lim\limits_{N
\to\infty}\frac{1}{N}\sum_{i=1}^N P_{\rm out}(i),
\end{align}
which can be written equivalently as 
\begin{eqnarray}\label{eq_nz_q_1a}
    \lim\limits_{N
\to\infty}\frac{1}{N}\sum_{i=1}^N B(i) - \lim\limits_{N
\to\infty}\frac{1}{N}\sum_{i=1}^N B(i-1)  = \bar P_{\rm in} - \bar P_{\rm out}.
\end{eqnarray}
Now, since $\bar P_{\rm d}<\bar P_{\rm in}$, and since  due to (\ref{eq_P_out_av})  $\bar P_{\rm out}\leq \bar P_{\rm d}$ holds, we have  $\bar P_{\rm out}<\bar P_{\rm in}$, which means that $\bar P_{\rm in} - \bar P_{\rm out}=\epsilon >0$. Replacing this in the right hand side of (\ref{eq_nz_q_1a}), and writing $\sum_{i=1}^N B(i)=\sum_{i=1}^N B(i-1)+B(N)$, where $B(0)=0$, we obtain the following
\begin{align}\label{eq_nz_q_1b}
   & \lim\limits_{N
\to\infty}\frac{1}{N}\sum_{i=1}^N B(i-1)    - \lim\limits_{N
\to\infty}\frac{1}{N}\sum_{i=1}^N B(i-1) \nonumber\\
&+ \lim\limits_{N
\to\infty}\frac{1}{N}  B(N)  = \epsilon.
\end{align}
The two sums in (\ref{eq_nz_q_1b}) cancel each other out, and we obtain the following identity for  $\bar P_{\rm d}=\bar P_{\rm lim}<\bar P_{\rm in}$
\begin{eqnarray}\label{q_qa}
    \lim\limits_{N
\to\infty}   B(N)  = \lim\limits_{N
\to\infty} N \epsilon =\infty.
\end{eqnarray}
Hence, when $\bar P_{\rm d}=\bar P_{\rm lim}<\bar P_{\rm in}$ holds, for any time slot $N\to\infty$, (\ref{q_qa}) holds. This is intuitive, since if in each time slot, on average, more energy is stored in an infinite storage battery than what is extracted from the battery, then, after $N\to\infty$ time slots, there must be infinite energy in the battery.  We now use (\ref{q_qa}) for proving the following proposition.
\begin{proposition}
When $\bar P_{\rm d}=\bar P_{\rm lim}<\bar P_{\rm in}$, there must be some time slot, denoted by $j$,   after which  for any $i>j$ the event  $P_{\rm out}(i)=\min\{B(i-1),P_{\rm d}(i)\}=B(i-1)$ does not occur, and for $i>j$, only the event  $P_{\rm out}(i)=\min\{B(i-1),P_{\rm d}(i)\}=P_{\rm d}(i)$ occurs. Moreover,  $j$ must satisfy
\begin{eqnarray}\label{eq_n_n_1}
    \lim_{N\to\infty}\frac{j}{N}=0.
\end{eqnarray}
\end{proposition}
\textit{Proof of Proposition 1:} We prove Proposition 1   by contradiction. Hence, we assume that we cannot find  a  time slot $j$, defined   in Proposition 1, since  $\min\{B(j-1),P_{\rm d}(j)\}=B(j-1)$ occurred for some large $j$ for which $\lim\limits_{N\to\infty} j/N\neq 0$ holds.
However,  if $\lim\limits_{N\to\infty} j/N\neq 0$ holds,  then   $j\to\infty$ must hold. Consequently, if $j\to\infty$ and $\min\{B(j-1),P_{\rm d}(j)\}=B(j-1)$ occurred for that $j$, then due to (\ref{q_qa})   we would get   the following   identity
\begin{align}\label{eq_app_1a}
 & \lim_{j\to\infty} \min\{B(j-1),P_{\rm d}(j)\} = \lim_{j\to\infty}  B(j-1)  =\infty.
\end{align}
However, since the power $P_{\rm d}(j)$ is finite, the expression in the    left hand side of (\ref{eq_app_1a}) must also satisfy
\begin{align}\label{eq_app_1b}
&\lim_{j\to\infty}   \min\{B(j-1),P_{\rm d}(j)\} \leq \lim_{j\to\infty}   P_{\rm d}(j)  <\infty,
\end{align}
which is a contradiction to (\ref{eq_app_1a}), i.e., we obtain that  for $j\to\infty$  both (\ref{eq_app_1a}) and  (\ref{eq_app_1b}) have to hold, which is impossible.   As a result, Proposition 1 must be true.

Now if Proposition 1 is true, then the number of time slots $\Delta$ for which $P_{\rm out}(i)\neq P_{\rm d}(i)$ holds, satisfies $\lim_{N\to\infty} \Delta/N=0$, and the number of time slots for which $P_{\rm out}(i)=P_{\rm d}(i)$ holds satisfies $\lim_{N\to\infty} (N-\Delta)/N=1$. Thereby,   $P_{\rm out}(i)=P_{\rm d}(i)$ holds practically always. This concludes the proof that $P_{\rm out}(i)=P_{\rm d}(i)$ holds practically always.

On the other hand, since $P_{\rm d}(i)$ and  $U(i)$, $\forall i$, are finite, and since $j$ is finite, we  
have
\begin{align}
&\lim_{N\to\infty}\frac{1}{N}\sum_{i=1}^j   P_{\rm d}(i)=0\label{elmsm_1}\\
&\lim_{N\to\infty}\frac{1}{N}\sum_{i=1}^j \min\{B(i-1), P_{\rm d}(i)\} =0 \label{elmsm_2} \\
&\lim_{N\to\infty}\frac{1}{N}\sum_{i=1}^j U(i)=0 .\label{elmsm_3}
\end{align}
From (\ref{elmsm_1})-(\ref{elmsm_3}), we obtain the following  
\begin{align}
   \bar P_{\rm out} &= \lim_{N\to\infty}\frac{1}{N}\sum_{i=1}^j \min\{B(i-1), P_{\rm d}(i)\}\nonumber\\
&+\lim_{N\to\infty}\frac{1}{N}\sum_{i=j+1}^N P_{\rm d}(i) =\bar P_{\rm d}. \label{eq_qwe} \\
    \bar U&= \lim_{N\to\infty}\frac{1}{N}\sum_{i=1}^j U(i)+ \lim_{N\to\infty}\frac{1}{N}\sum_{i=j+1}^N U(i)\nonumber\\
& = \lim_{N\to\infty}\frac{1}{N}\sum_{i=j+1}^N U(i),
\end{align}
i.e., $\bar P_{\rm out}=\bar P_{\rm d}$ and only the codewords after the $j$-th slot contribute to the average utility function, the contribution of the other codewords is negligible. Since for $i>j$, $P_{\rm out}(i)=P_{\rm d}(i)$ holds always, we obtain that only the codewords for which $P_{\rm out}(i)=P_{\rm d}(i)$ holds contribute to the average utility function $\bar U$ and the other codewords have  negligible contributions to $\bar U$.
 This completes the proof of Part I.

\subsubsection{ Part II ($\bar P_{\rm d}=\bar P_{\rm in}\leq \bar P_{\rm lim}$)} 
If $\bar P_{\rm d}= \bar P_{\rm in} $,  the following must hold
\begin{align}\label{usl_d1}
 \bar P_{\rm out}&= \lim_{N\to\infty}\frac{1}{N}\sum_{i=1}^N \min\{B(i-1),P_{\rm d}(i)\} \nonumber\\
&= \lim_{N\to\infty}\frac{1}{N}\sum_{i=1}^N P_{\rm d}(i)= \bar P_{\rm d} = \bar P_{\rm in}  .
\end{align}
We prove the above claim  by contradiction. Assume that $\bar P_{\rm d}= \bar P_{\rm in} $ holds, however, (\ref{usl_d1}) does not hold  and  
 \begin{align}\label{usl_d2}
  \bar P_{\rm out}&=  \lim_{N\to\infty}\frac{1}{N}\sum_{i=1}^N \min\{B(i-1),P_{\rm d}(i)\}\nonumber\\
&<\lim_{N\to\infty}\frac{1}{N}\sum_{i=1}^N P_{\rm d}(i)=  \bar P_{\rm d} = \bar P_{\rm in} 
\end{align}
holds instead. However, since $\bar P_{\rm out}<\bar P_{\rm in}$ holds, according to the  proof in Part I,  $\bar P_{\rm out}$ has to be given by (\ref{eq_qwe}). Thereby, we obtain a contradiction that both   (\ref{eq_qwe}) and (\ref{usl_d2}) must hold. Due to this contradiction, (\ref{usl_d2}) cannot hold and (\ref{usl_d1}) must hold instead. This concludes the proof of  (\ref{usl_d1}).

Now, there are only two cases for which (\ref{usl_d1}) can   hold. The first case is when   the number of slots for which $P_{\rm out}(i)=B(i-1)$ holds, denoted by $\Delta$, is negligible compared to the number of slots for which $P_{\rm out}(i)=P_{\rm d}(i)$ holds, i.e., $\Delta$ is such that $\lim_{N\to\infty}{\Delta}/N=0$ holds. As a result, the  codewords for which  $P_{\rm out}(i)=B(i-1)$ occurs have negligible effect on the average utility function $\bar U$ compared to the codewords for which $P_{\rm out}(i)=P_{\rm d}(i)$ occurs. This can be proved as follows. Put all time slots $i$ for which $P_{\rm out}(i)=B(i-1)$ occurs into set $\mathcal{I}$ and the rest of time slots for which $P_{\rm out}(i)=P_{\rm d}(i)$ occurs into set $\bar{\mathcal{I}}$ . Then, let $U_{\max}=\max_{i\in\mathcal{I}} U(i)$. Since $U_{\max}$ is finite it follows that the contribution of the codewords with powers $P_{\rm out}(i)=B(i-1)$ to the average utility function is 
\begin{align}
 &   \lim_{N\to\infty}\frac{1}{N}\sum_{i\in \mathcal{I}} U(i)\leq \lim_{N\to\infty}\frac{1}{N}\sum_{i\in \mathcal{I}} U_{\max} \nonumber\\
& = \lim_{N\to\infty}\frac{\Delta}{N} \times U_{\max}=0.
\end{align}
The second case is when the number of slots for which $P_{\rm out}(i)=B(i-1)$ holds,  $\Delta$, is not negligible compared to the number of slots for which $P_{\rm out}(i)=P_{\rm d}(i)$ holds, i.e., $\Delta$ is such that $\lim_{N\to\infty}{\Delta}/N>0$ holds. Nevertheless, we can prove  when $P_{\rm out}(i)=B(i-1)$ occurs, the difference between $B(i-1)$ and  $P_{\rm d}(i)$ must be so small that its effect on $\bar U$ is negligible. As a result of this negligible effect on $\bar U$, we can assume that  $P_{\rm out}(i)=B(i-1)=   P_{\rm d}(i)$ holds for practically all time slots.  This is proven in the following.

Since (\ref{usl_d1}) holds,  using the sets $\mathcal{I}$ and  $\bar{\mathcal{I}}$, we can rewrite  (\ref{usl_d1}) as
\begin{align}\label{usl_dsd1}
\lim_{N\to\infty}\frac{1}{N}\sum_{i=1}^N P_{\rm d}(i) &= \lim_{N\to\infty}\frac{1}{N}\sum_{i\in\mathcal{I}}^N  B(i-1)  \nonumber\\
&+ \lim_{N\to\infty}\frac{1}{N}\sum_{i\in\bar{\mathcal{I}}}^N  P_{\rm d}(i). 
\end{align}
Substracting the right hands side of (\ref{usl_dsd1}) from both sides of  (\ref{usl_dsd1}), we obtain that 
\begin{align}\label{eq_sdf}
 \lim_{N\to\infty}\frac{1}{N}\sum_{i\in\mathcal{I}}^N \big(P_{\rm d}(i)- B(i-1)\big) =0  
\end{align}
must hold.
Now, since for $i\in\mathcal{I}$, $P_{\rm d}(i)>  B(i-1) $ holds, we can conclude that    (\ref{eq_sdf}) can hold if and only if in almost all time slots  $i\in\mathcal{I}$, the difference between $P_{\rm d}(i)$  and $B(i-1) $ is negligible. Or, more precisely, for $i\in\mathcal{I}$, the average  of the difference 
\begin{align}\label{eq_dfrd}
  \epsilon(i)= P_{\rm d}(i)- B(i-1)>0 , \;\; i\in\mathcal{I},
\end{align}
    satisfies
\begin{align}\label{eq_sdfdf}
 \lim_{N\to\infty}\frac{1}{N}\sum_{i\in\mathcal{I}}^N \epsilon(i) =0. 
\end{align}
In the following, we prove that  the events in which $P_{\rm out}(i)=B(i-1)$ occurs have negligible effect on  $\bar U$. To this end, let us define $\bar U_{\rm d}$ as the average utility function obtained when $P_{\rm out}(i)=P_{\rm d}(i)$, $\forall i$. Then, we can write $\bar U_{\rm d}$  and $\bar U$ 
\begin{align}
  \bar U_{\rm d}&= \lim_{N\to\infty}\frac{1}{N}\sum_{i=1}^N  U(P_{\rm d}(i))\nonumber\\
&= 
\lim_{N\to\infty}\frac{1}{N}\sum_{i\in\mathcal{I}}  U(P_{\rm d}(i)) +  \lim_{N\to\infty}\frac{1}{N}\sum_{i\in\bar{\mathcal{I}}}  U(P_{\rm d}(i))  \label{eq_hgj} \\
 \bar U&= \lim_{N\to\infty}\frac{1}{N}\sum_{i=1}^N  U(P_{\rm out}(i)) \nonumber\\
&= 
\lim_{N\to\infty}\frac{1}{N}\sum_{i\in\mathcal{I}}  U(B(i-1)) +  \lim_{N\to\infty}\frac{1}{N}\sum_{i\in\bar{\mathcal{I}}}  U(P_{\rm d}(i))  .
\end{align}
Now, using (\ref{eq_dfrd}), we can obtain $P_{\rm d}(i)$ for $i\in\mathcal{I}$ as  $P_{\rm d}(i)=  B(i-1) +\epsilon(i)$. Inserting this into  (\ref{eq_hgj})  we can obtain  $\bar U_{\rm d}$ as
\begin{align}\label{eq_hgjfsd}
 & \bar U_{\rm d}
 =\nonumber\\
&\lim_{N\to\infty}\frac{1}{N}\sum_{i\in\mathcal{I}}  U(B(i-1) +\epsilon(i)) +  \lim_{N\to\infty}\frac{1}{N}\sum_{i\in\bar{\mathcal{I}}}  U(P_{\rm d}(i)) \nonumber\\
&\stackrel{(a)}{=} \lim_{N\to\infty}\frac{1}{N}\sum_{i\in\mathcal{I}}  U(B(i-1) ) +  \lim_{N\to\infty}\frac{1}{N}\sum_{i\in\bar{\mathcal{I}}}  U(P_{\rm d}(i)) 
=\bar U ,
\end{align}
 where $(a)$ follows  from the fourth property of   $\bar U$ given in Definition~\ref{def_1}.

Hence,  also for the second part  when $\bar P_{\rm d}=\bar P_{\rm in}\leq \bar P_{\rm lim}$ holds, we can conclude that $P_{\rm out}(i)= P_{\rm d}(i)$   holds  for  practically all time slots.
This completes the proof of Lemma~\ref{lemma_1}.

\subsection{Proof of Theorem~\ref{theo_3}}\label{app_3}
Following the same procedure as for the proofs in Appendix \ref{app_1},  with  $\bar P_{\rm d}$ adjusted to $\bar P_{\rm d}=\min\{\bar P_{\rm lim},\bar P_{\rm in}\}$,  we obtain that $P_{\rm out}(i)=P_{\rm d}(i)$   holds for practically all time slots when $N\to\infty$. Now, for each time slot $i$ for which $P_{\rm out}(i)=P_{\rm d}(i)$ holds, $P_{{\rm out},k}(i)=P_{{\rm d},k}(i)$, $\forall k$,   also holds. Therefore, it follows that $P_{{\rm out},k}(i)=P_{{\rm d},k}(i)$, $\forall k$,   holds for practically all time slots  when $N\to\infty$. Consequently, following the method in the proof of Theorem~\ref{theo_1}, we can write (\ref{MPR1-EH-bc}) as  (\ref{MPR1_non-EH-bc}) with appropriately adjusted $P_{\rm lim,non-EH}=\min\{\bar P_{\rm lim},\bar P_{\rm in}\}$ as explained in Theorem~\ref{theo_3}.

\subsection{Proof of Theorem~\ref{theo_4}}\label{app_4}
In this proof,   set $\mathcal{A}$ comprises  the indices of the nodes with $\bar P_{{\rm lim},k}<\bar P_{{\rm in},k} $ and    set $\bar{\mathcal{A}}$ comprises  the indices of the nodes with   $\bar P_{{\rm lim},k}\geq \bar P_{{\rm in},k} $.
We first prove that for the nodes   $k\in  \mathcal{A} $,  the number of time slots for which $P_{{\rm out},k}(i) \neq P_{{\rm d},k}(i) $ holds is  negligible compared to the number of time slots with $P_{{\rm out},k}(i)= P_{{\rm d},k}(i)$. To this end, we use the
proof in Appendix~\ref{app_1}, where it is shown that an individual   node $k\in \mathcal{A}$, after some finite number of time slots $j_k$,   transmits with power $P_{{\rm out},k}(i)=P_{{\rm d},k}(i)$, $\forall i>j_k$, where $\lim_{N\to\infty} j_k/N=0$. Now, let $j=\max_k\{j_k\}$. Then, $\lim_{N\to\infty} j/N=0$ holds. Furthermore, after this $j$-th time slot, all of the nodes    $k\in \mathcal{A}$ transmit with power  $P_{{\rm out},k}(i)=P_{{\rm d},k}(i)$, $\forall i>j$, i.e., these nodes transmit with power  $P_{{\rm out},k}(i)=P_{{\rm d},k}(i)$ for practically all time slots. This completes the proof for the  nodes in the set  set $\mathcal{A}$.
 Now, we are only left to prove that for the nodes   $k\in \bar{\mathcal{A}}$, the number of time slots in which $P_{{\rm out},k}(i)\neq P_{{\rm d},k}(i) $ holds is  negligible compared to the number of time slots with $P_{{\rm out},k}(i)= P_{{\rm d},k}(i)$. To this end,  for each  node  $k\in \bar{\mathcal{A}}$,  let us create a set $\mathcal{I}_k$ in which we put the time slots $i$ for which $P_{{\rm out},k}(i)\neq P_{{\rm d},k}(i) $ holds. Furthermore, let us create the set $\mathcal{I}$ in which we put the time slots $i$ for which  $P_{{\rm out},k}(i)\neq P_{{\rm d},k}(i) $ holds for at least one of the  nodes $k\in \bar{\mathcal{A}}$, i.e.,  $\mathcal{I}$ is the union of all $\mathcal{I}_k$ for $k\in \bar{\mathcal{A}}$. Thereby, the cardinality of $\mathcal{I}$ is upper bounded by the sum of the cardinalities of $\mathcal{I}_k$, $\forall k\in \bar{\mathcal{A}}$, i.e., the cardinality of $\mathcal{I}$, denoted by  $\Delta$, is upper bounded as 
\begin{eqnarray}\label{eq_kk_1}
    \Delta=|\mathcal{I}|\leq \sum_{k\in \bar{\mathcal{A}}}  |\mathcal{I}_k|.
\end{eqnarray}
According to the proof in Appendix~\ref{app_1}, if $\bar P_{{\rm d},k}=\bar P_{{\rm in},k}$ holds,  for any individual node $k$, $P_{{\rm out},k}(i)= P_{{\rm d},k}(i)$  hold for practically all time slots. Hence, let us adopt $\bar P_{{\rm d},k}=\bar P_{{\rm in},k}$, $\forall k\in \mathcal{\bar A} $. Now, since $|\mathcal{I}_k|$ is the number of time slots in which  $P_{{\rm out},k}(i) \neq P_{{\rm d},k}(i) $ holds for node $k\in \mathcal{\bar A}$,    according to the proof in Appendix~\ref{app_1}, $|\mathcal{I}_k|$  satisfies
\begin{eqnarray}\label{eq_kk_2}
    \lim_{N\to\infty} |\mathcal{I}_k|/N=0\;, \forall k\in \bar{\mathcal{A}}.
\end{eqnarray} 
%Note that Theorem~2 also proves that when $\bar P_{{\rm d},k}=\bar P_{{\rm in},k}$ holds, then  $\bar P_{{\rm out},k}$ reaches its maximum value which is $\bar P_{{\rm out},k}=\bar P_{{\rm in},k}$. Since $U(i)$ is a monotonic function in terms of $P_{{\rm out},k}(i)$, $\forall i,k$, it follows that $\bar U$ reaches its maximal value when  $\bar P_{{\rm out},k}$, $\forall k \in \bar{\mathcal{A}}$, reaches also its maximal value. Hence, the maximum of $\bar U$ can be achieved when  $\bar P_{{\rm d},k}=\bar P_{{\rm in},k}$ holds $\forall k \in\bar{\mathcal{A}}$, since then   $\bar P_{{\rm out},k}$, $\forall k \in\bar{\mathcal{A}}$, reach their maximal value.
 Now,  combining (\ref{eq_kk_1}) and (\ref{eq_kk_2}) we find that the cardinality of $\mathcal I$ satisfies
\begin{eqnarray}\label{eq_kk_3}
    \lim_{N\to\infty} \Delta/N\leq \sum_{k\in \bar{\mathcal{A}}}  \lim_{N\to\infty} |\mathcal{I}_k|/N=0.
\end{eqnarray} 
when $|\mathcal{\bar A}|<\infty$.
 Therefore, the cumulative effect  of $P_{{\rm out},k}(i)\neq P_{{\rm d},k}(i) $  on $\bar U$ from the  nodes in $\mathcal{A}$ and   $\mathcal{\bar A}$    is negligible since $\lim_{N\to\infty} (j+\Delta)/N=0$ holds.  Therefore,   following the method in the proof of Theorem~\ref{theo_1},  we can write (\ref{MPR1-EH-mac}) as  (\ref{MPR1_non-EH-mac}) with appropriately adjusted $P_{\rm lim,non-EH}$ as explained in Theorem~\ref{theo_4}. This completes the proof.

\subsection{Proof of Theorem~\ref{theo_5}}\label{app_5}
The proof for the  general EH network is identical to that of the multiple-access EH network given in Appendix \ref{app_4}, however, the   powers $P_{{\rm out},k}(i)$ and $P_{{\rm d},k}(i)$ now have different meanings and  are given by (\ref{e_1}) and (\ref{e_2}), respectively. In particular, $P_{{\rm out},k}(i)$ and $P_{{\rm d},k}(i)$ now are the total transmit and the total desired powers in time slot $i$ of the $k$-th EH transmit node to all receiving nodes. Hence, by following the same procedure as for the proof in Appendix \ref{app_4}, we can prove that, if for the   nodes $k\in\mathcal{A}$ for which $\bar P_{{\rm lim},k}<\bar P_{{\rm in},k} $ holds  and the  nodes $ k\in\bar{\mathcal{A}}$ for which  $\bar P_{{\rm lim},k}\geq \bar P_{{\rm in},k} $ holds,  $\bar P_{{\rm d},k}$ is set to $\bar P_{{\rm d},k}=\bar P_{{\rm lim},k}$ and $\bar P_{{\rm d},k}=\bar P_{{\rm in},k}$, respectively, we obtain that $P_{{\rm out},k}(i) = P_{{\rm d},k}(i) $   holds for practically all time slots. As a result,   $P_{{\rm out},k\to j}(i) =P_{{\rm d},k\to j}(i)$ also  holds  $\forall i,k,j$. Hence,  Theorem~\ref{theo_5} follows.
\end{appendix}

%%%%%%%%%%%%%%%%%%%%%%%%%%%%%%%%%%%%%%%%%%%
\bibliography{litdab}
\bibliographystyle{IEEEtran}
% that's all folks

\begin{IEEEbiography}[{\includegraphics[width=1.1in,height=1.8in,clip,keepaspectratio]{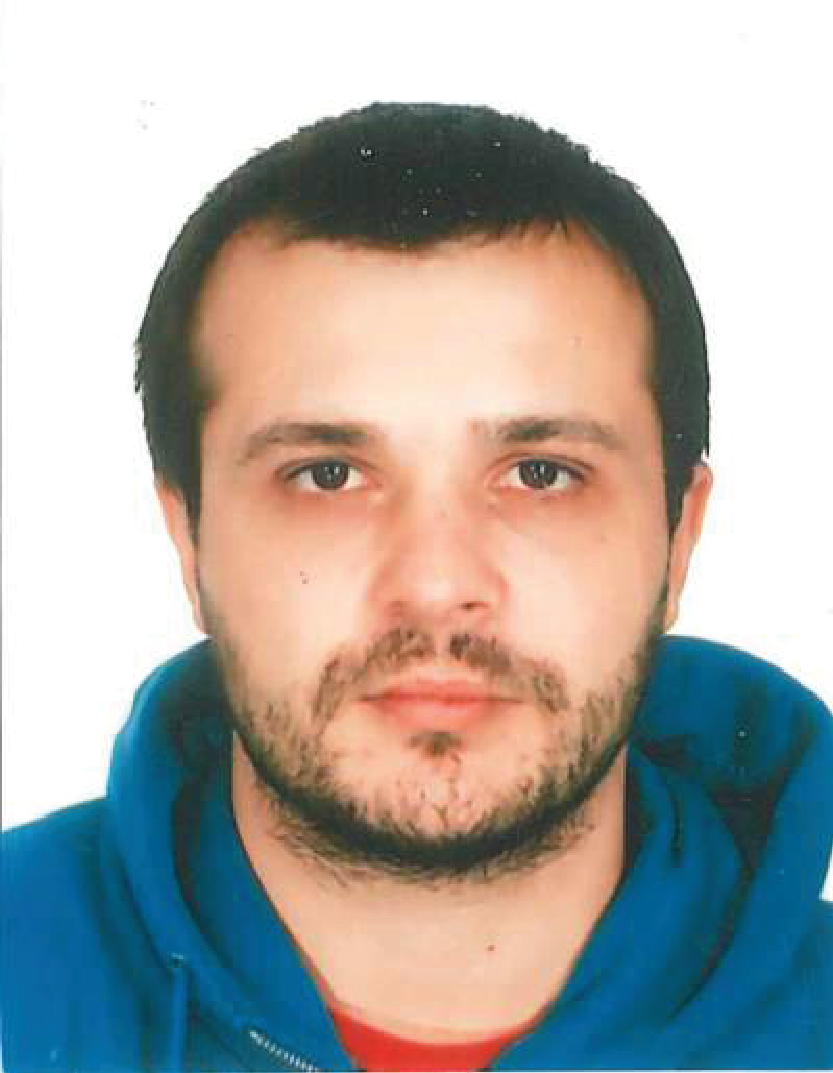}}]{Nikola Zlatanov (S'06, M'15)}
 was born in Macedonia. He received the Dipl.Ing. and Master degree in electrical engineering from Ss. Cyril and Methodius University, Skopje, Macedonia in 2007 and 2010, respectively, and his PhD degree from the University of British Columbia (UBC) in Vancouver, Canada in 2015. He is currently a Lecturer (Assistant Professor) in the Department of Electrical and Computer Systems Engineering   at Monash University in Melbourne, Australia. His current research interests include wireless communications and information theory.
Dr. Zlatanov received several scholarships/awards for his work including UBC's Four Year Doctoral Fellowship in 2010, UBC's Killam Doctoral Scholarship and Macedonia's Young Scientist of the Year in 2011, the Vanier Canada Graduate Scholarship in 2012, best journal paper award from the German Information Technology Society (ITG) in 2014, and best conference paper award at ICNC in 2016. Dr. Zlatanov serves as an Editor of IEEE Communications Letters. He has been a TPC member of various conferences, including  Globecom,   ICC, VTC, and ISWCS.
\end{IEEEbiography}

\begin{IEEEbiography}[{\includegraphics[width=1.1in,height=1.45in,clip,keepaspectratio]{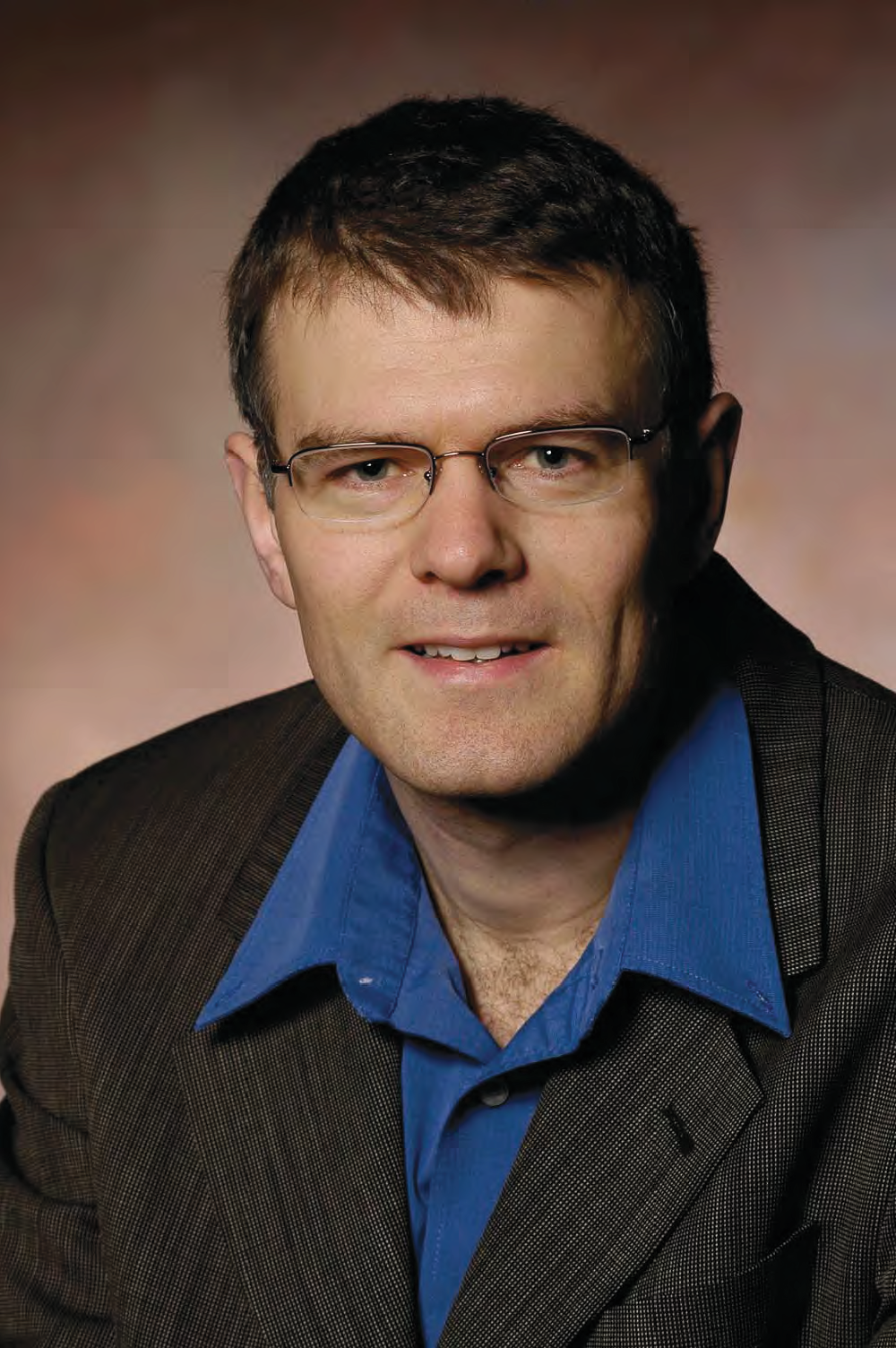}}]{Robert Schober (S'98, M'01, SM'08, F'10)}
  was born in Neuendettelsau, Germany, in 1971. He received the Diplom (Univ.) and the Ph.D. degrees in electrical engineering from the University of Erlangen-Nuermberg in 1997 and 2000, respectively. From May 2001 to April 2002 he was a Postdoctoral Fellow at the University of Toronto, Canada, sponsored by the German Academic Exchange Service (DAAD). From 2002 to 2011, he was a Professor and Canada Research Chair at the University of British Columbia (UBC), Vancouver, Canada. Since January 2012 he is an Alexander von Humboldt Professor and the Chair for Digital Communication at the Friedrich Alexander University (FAU), Erlangen, Germany. His research interests fall into the broad areas of Communication Theory, Wireless Communications, and Statistical Signal Processing.

Dr. Schober received several awards for his work including the 2002 Heinz Maier–Leibnitz Award of the German Science Foundation (DFG), the 2004 Innovations Award of the Vodafone Foundation for Research in Mobile Communications, the 2006 UBC Killam Research Prize, the 2007 Wilhelm Friedrich Bessel Research Award of the Alexander von Humboldt Foundation, the 2008 Charles McDowell Award for Excellence in Research from UBC, a 2011 Alexander von Humboldt Professorship, and a 2012 NSERC E.W.R. Steacie Fellowship. In addition, he has received several best paper awards for his research. Dr. Schober is a Fellow of the Canadian Academy of Engineering and a Fellow of the Engineering Institute of Canada. From 2012 to 2015, he served as Editor-in-Chief of the IEEE Transactions on Communications and since 2014, he is the Chair of the Steering Committee of the IEEE Transactions on Molecular, Biological and Multiscale Communication. Furthermore, he is a Member at Large of the Board of Governors of the IEEE Communications Society.
\end{IEEEbiography}

\begin{IEEEbiography}[{\includegraphics[width=1.1in,height=1.8in,clip,keepaspectratio]{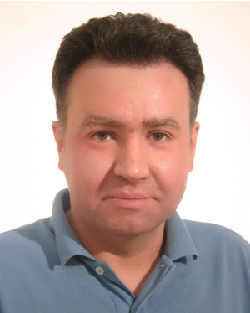}}]{Zoran Hadzi-Velkov  (M'97, SM'11)}
 received Dipl.-Ing. in electrical engineering (honors), Magister Ing. in communications engineering (honors), and Ph.D. in technical sciences from the Ss. Cyril and Methodius University in Skopje, Macedonia, in 1996, 2000, and 2003, respectively. He is currently a Professor of telecommunications at his alma mater. During 2001 and 2002, he was a visiting scholar at the IBM Watson Research Center, New York, USA. Between 2012 and 2014, Dr. Hadzi-Velkov was a visiting professor at the Institute for Digital Communications, University of Erlangen-Nuremberg in Germany. He received the Alexander von Humboldt fellowship for experienced researchers in 2012, and the annual best scientist award from Ss. Cyril and Methodius University in 2014. Between 2012 and 2015, he was the Chair of the Macedonian Chapter of IEEE Communications Society. He has served on the technical program committees of numerous international conferences, including IEEE ICC 2013, IEEE ICC 2014, IEEE ICC 2015, IEEE GLOBECOM 2015, and IEEE GLOBECOM 2016. Since 2012, Dr. Hadzi-Velkov served as an Editor for IEEE COMMUNICATIONS LETTERS. His research interests are in the broad area of wireless communications, with particular emphasis on cooperative communications, green communications, and energy harvesting communications. 
\end{IEEEbiography}

\end{document}